\documentclass[twocolumn]{aastex63}

\usepackage{amsmath}
\usepackage{multirow}
\usepackage{textcomp}

\newcommand{\mic}{\textmu m}

\newcommand{\Rs}{R$_\odot$}
\newcommand{\Bs}{B$_\odot$}

\newcommand{\revis}[1]{#1}
\newcommand{\revisb}[1]{\textcolor{blue}{#1}}

\shorttitle{Airborne Infrared Spectrometer}
\shortauthors{Samra et al.}

\begin{document}

\title{The Airborne Infrared Spectrometer: \\Development, Characterization, and the 21 August 2017 Eclipse Observation}
\correspondingauthor{Jenna Samra}
\email{jsamra@cfa.harvard.edu}

\author{Jenna E. Samra}
\affiliation{Smithsonian Astrophysical Observatory, 60 Garden Street, Cambridge, MA, 02138, USA}

\author{Vanessa Marquez}
\affiliation{Smithsonian Astrophysical Observatory, 60 Garden Street, Cambridge, MA, 02138, USA}

\author{Peter Cheimets}
\affiliation{Smithsonian Astrophysical Observatory, 60 Garden Street, Cambridge, MA, 02138, USA}

\author{Edward E. DeLuca}
\affiliation{Smithsonian Astrophysical Observatory, 60 Garden Street, Cambridge, MA, 02138, USA}

\author{Leon Golub}
\affiliation{Smithsonian Astrophysical Observatory, 60 Garden Street, Cambridge, MA, 02138, USA}

\author{James W. Hannigan}
\affiliation{Atmospheric Chemistry Observations and Modeling, National Center for Atmospheric Research,
P.O. Box 3000, Boulder, CO 80307, USA}

\author{Chad A. Madsen}
\affiliation{Smithsonian Astrophysical Observatory, 60 Garden Street, Cambridge, MA, 02138, USA}

\author{Alisha Vira}
\affiliation{Georgia Institute of Technology, 225 North Avenue, Atlanta, GA 30332, USA}

\author{Arn Adams}
\affiliation{IRCameras, 30 S. Calle Cesar Chavez, Santa Barbara, CA 93103, USA}

\begin{abstract}
On August 21, 2017, the Airborne Infrared Spectrometer (AIR-Spec) observed the total solar eclipse at an altitude of 14 km from aboard the NSF/NCAR Gulfstream V research aircraft. The instrument successfully observed the five coronal emission lines that it was designed to measure: \ion{Si}{10}~1.431~\mic, \ion{S}{11} 1.921~\mic, \ion{Fe}{9} 2.853~\mic, \ion{Mg}{8} 3.028~\mic, and \ion{Si}{9} 3.935~\mic. Characterizing these magnetically sensitive emission lines is an important first step in designing future instruments to monitor the coronal magnetic field, which drives space weather events as well as coronal heating, structure, and dynamics. The AIR-Spec instrument includes an image stabilization system, feed telescope, grating spectrometer, and slit-jaw imager. This paper details the instrument design, optical alignment method, image processing, and data calibration approach. The eclipse observations are described and the available data are summarized.\\
\end{abstract}



\section{Introduction} 
\label{sec:introduction}

Observations of infrared (IR) solar coronal emission lines are currently of significant interest, motivated by the need to directly measure coronal magnetic fields \citep{Judge2001,Penn2014} and by the advent of new instrumentation, especially the Daniel K. Inouye Solar Telescope (DKIST, \citealp{Tritschler2016,Rimmele2020}). Many IR lines are strongly photoexcited and are therefore promising for accurately measuring Doppler shifts and non-thermal line widths at significant heights above the limb \citep{DelZanna2018}. In addition, observations of these lines are needed to inform the design of new instrumentation for coronal magnetometry.

In their coronal magnetometry feasibility study, \citet{Judge2001} find that the two most promising methods for observing the coronal magnetic field are measuring the Zeeman and Hanle effects in polarized emission lines. The Zeeman effect modulates the circular polarization and provides a means of measuring the line-of-sight field magnitude, while the Hanle effect is encoded in the linear polarization and gives a measurement of the plane-of-sky field direction \citep{Casini1999, Lin2000, Judge2001}. For sufficient sensitivity, circular polarization measurements require emission lines at visible or IR wavelengths. Magnetically-induced Zeeman splitting varies with wavelength as $\lambda^2$, while thermal broadening goes as $\lambda$.  In addition, the effects of instrumental and atmospheric scattering are less significant at longer wavelengths.

\begin{figure}[!htp]
	\centering
	\includegraphics[angle=0,width=0.92\linewidth]{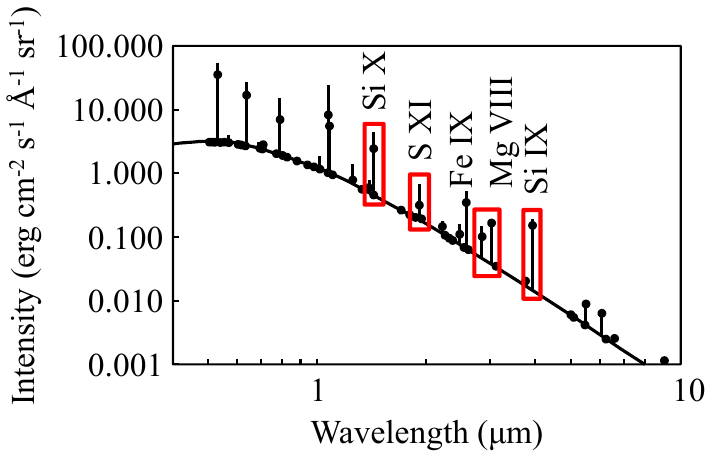}
	\caption{Predicted spectrum of M1 lines at 1.1 \Rs\ \citep{Judge1998}. Black dots are intensity values for collisional de-excitation only, straight lines include radiative contributions. The smooth baseline is the coronal continuum produced by Thomson scattering by free electrons. Lines outlined in red were measured by AIR-Spec during the 2017 eclipse.}
	\label{fig:judge-spectrum}
\end{figure}

In the visible and infrared, strong coronal emission lines are formed by forbidden magnetic dipole (M1) transitions. A modeled spectrum (a reproduction of Figure 2 from \citealp{Judge1998}) is shown in Figure~\ref{fig:judge-spectrum}. Historically, the corona has been observed infrequently above 1~\mic\ \citep{DelZanna2018}, and these lines are not as well-understood as their visible and near-IR counterparts.  \revis{Exploration of this wavelength region has become more affordable with the availability of high-sensitivity, large-format, commercial off-the-shelf (COTS) detectors.}

Based on the modeled intensities in Figure \ref{fig:judge-spectrum}, \citet{Judge2001} list the most promising candidates for magnetic field measurements (reproduced in Table \ref{tab:M1-lines}). The lines near the center of the table are good candidates for making magnetic field measurements because they balance the short wavelength advantages of high line intensity and low thermal emission with the long-wavelength benefits of high sensitivity to magnetic field and low scattered light. However, these candidate lines must be characterized before spectro-polarimeters can be designed around the most useful ones. At the time of the 2017 total solar eclipse, five of the promising lines in Table \ref{tab:M1-lines} had never been observed: \ion{Si}{9} 2.584~\mic, \ion{Fe}{9} 2.855~\mic, and the three lines with $\lambda$\;$>$\;5 \mic\ \citep{DelZanna2018}.  Most of the IR lines that had been measured were not well-characterized spatially or temporally (e.g. \citealp{Olsen1971,Kuhn1996,Judge2002}). 

\begin{deluxetable}{lccclllcc}[!htp]
\tablecaption{Candidate M1 lines for coronal magnetic field measurements \citep{Judge2001}.}
\label{tab:M1-lines}
\tablehead{\colhead{\multirow{2}{*}{Ion}} & \colhead{$\lambda$} & \colhead{$\log T_e$} & \colhead{$F_E$} & \colhead{$F_{scat}$} & \colhead{$F_{th}$} & \colhead{$T$}\\
		& \colhead{\mic}& \colhead{K} & \multicolumn{3}{c}{------ ph cm$^{-2}$ s$^{-1}$ ------} &\colhead{\%}}
\startdata
\ion{Fe}{14} & 0.530 & 6.30 & 2.57+8 & 2.78+8 & 8.22--32 & 100 \\
\ion{Fe}{10} & 0.637 & 6.03 & 1.77+8 & 1.37+8 & 3.96--24 & 100 \\
\ion{Fe}{11} & 0.789 & 6.10 & 1.48+8 & 5.48+7 & 7.21--17 & 100 \\
\ion{Fe}{13} & 1.075 & 6.22 & 4.91+8 & 1.26+7 & 7.27--9 & 100 \\
\bf\ion{Si}{10} & \bf1.430 & \bf6.13 & \bf1.57+8 & \bf2.88+6 & \bf1.85--3 & \bf\phantom{0}50 \\
\bf\ion{S}{11} & \bf1.920 & \bf6.25 & \bf3.59+7 & \bf5.85+5 & \bf2.21+1 & \bf\phantom{0}50 \\
\ion{Si}{9} & 2.584 & 6.04 & 6.07+7 & 1.11+5 & 2.01+4 & \phantom{00}0 \\
\bf\ion{Fe}{9} & \bf2.855 & \bf5.94 & \bf1.60+7 & \bf6.31+4 & \bf1.23+5 & \bf\phantom{0}60 \\
\bf\ion{Mg}{8} & \bf3.027 & \bf5.92 & \bf2.73+7 & \bf4.51+4 & \bf3.27+5 & \bf100 \\
\bf\ion{Si}{9} & \bf3.935 & \bf6.04 & \bf5.67+7 & \bf1.01+4 & \bf1.17+7 & \bf\phantom{0}60 \\
\ion{Mg}{7} & 5.502 & 5.80 & 3.20+6 & 1.42+3 & 2.80+8 & \phantom{0}10 \\
\ion{Fe}{11} & 6.081 & 6.10 & 2.61+6 & 7.88+2 & 5.62+8 & \phantom{0}50 \\
\ion{Mg}{7} & 9.031 & 5.80 & 9.53+5 & 7.66+1 & 3.79+9 & \phantom{0}90 \\
\enddata
\tablecomments{From left to right, the columns list the ion, transition wavelength, formation temperature in ionization equilibrium, emission line flux, scattered light, thermal emission, and atmospheric transmission at 8 km. The boldface rows correspond to transitions measured by AIR-Spec during the 2017 eclipse.}
\end{deluxetable}

This paper describes the Airborne Infrared Spectrometer (AIR-Spec), a new instrument that was developed to identify and characterize magnetically sensitive infrared coronal lines and assess their suitability for future ground-based, airborne, and space-based spectro-polarimetic observation. The lines targeted for observation are highlighted in Figure \ref{fig:judge-spectrum} and Table \ref{tab:M1-lines}.  Section~\ref{sec:instrument} describes the science goals and the instrument design, implementation, and deployment. Section \ref{sec:data} summarizes the observations, details the data processing and calibration schemes, describes the available data, and summarizes the lessons learned.

\section{The Airborne Infrared Spectrometer} \label{sec:instrument}
AIR-Spec is a \revis{slit spectrometer} that was designed to search for infrared emission lines of \ion{Si}{10}, \ion{S}{11}, \ion{Fe}{9}, \ion{Mg}{8}, and \ion{Si}{9} in the solar corona during the total solar eclipse on August 21, 2017. AIR-Spec was funded by a Major Research Instrumentation grant from the National Science Foundation (NSF) with cost-sharing by Smithsonian Institution. It observed the eclipse from the NSF/NCAR Gulfstream V High-performance Instrumented Airborne Platform for Environmental Research (GV HIAPER).

\subsection{Goals and Specifications}

The upper half of Table \ref{tab:perf-req} lists the predicted rest wavelengths, intensities, and linewidths of the five AIR-Spec lines. In addition to identifying its target lines, AIR-Spec was designed to characterize line emission as a function of solar conditions and radius, providing information on the radiative excitation of each line \citep{Habbal2011}, and to search for time-varying Doppler velocities in the lines, including high frequency velocity oscillations which are thought to be the signatures of waves or flows \citep{Tomczyk2007,DePontieu2010}. The instrument performance, shown in the lower half of Table \ref{tab:perf-req}, is sufficient to address these goals.

\begin{deluxetable*}{lccccc}[!htp]
	\tablecaption{Predicted emission line properties and measured instrument performance.}
	\label{tab:perf-req}
	\tablehead{& \colhead{\ion{Si}{10}} & \colhead{\ion{S}{11}} & \colhead{\ion{Fe}{9}} & \colhead{\ion{Mg}{8}} & \colhead{\ion{Si}{9}}}
	\startdata
	Rest wavelength\tablenotemark{a} (\mic) & 1.43 & 1.92 & 2.86 & 3.03 & 3.93 \\
	Thermal linewidth\tablenotemark{b} (\AA) & 2.3 & 3.1 & 4.6 & 4.9 & 6.3   \\
	Intensity\tablenotemark{c} at 1.1 \Rs & 39 & 8.8 & 4.0 & 6.7 & 13 \\
	($10^{11}$ ph s$^{-1}$ cm$^{-2}$ sr$^{-1}$) & 55 & 5.1 & 7.2 & 14 & 12 \\
	\tableline
	Spectral range (\mic) & 1.42 -- 1.54 & 1.87 -- 1.99 & 2.83 -- 3.07 & 2.83 -- 3.07 & 3.75 -- 3.98  \\
	Spectral dispersion (\AA/pixel) & 1.19 & 1.17 & 2.37 & 2.37 & 2.33  \\
	Spatial FOV (\Rs) & 1.55 & 1.55 & 1.55 & 1.55 & 1.55  \\
	Spatial sampling (arcsec/pixel) & 2.31 &  2.31 &  2.31 &  2.31 &  2.31 \\
	Spectral resolution (\AA) & 7.5 & 7.5 & 15 & 15 & 15 \\
	Spatial resolution (arcsec) & 11 & 13 & 11 & 11 & 13  \\		
	SNR\tablenotemark{d} of line center & 140 & 10 & 7 & 7 & 12 \\
	\enddata
	\tablenotetext{a}{Rest wavelengths were predicted by \citet{Judge1998}.}
	\tablenotetext{b}{Thermal line widths were extrapolated from measurements of the 530.3 nm \ion{Fe}{14} \citep{Contesse2004}.}
	\tablenotetext{c}{Original intensities (top row) were predicted by \citet{Judge1998}. Updated atomic data were used in new intensity estimates (bottom row) by \citet{DelZanna2018}.}
	\tablenotetext{d}{SNR is specified for a 30 second exposure over 35 arcsec at the limb.}
\end{deluxetable*}

The spectral range extends from each predicted central wavelength $\pm$200 \AA, allowing for the 0.1\% wavelength uncertainty \citep{Judge1998} plus a 500\% margin. The spectral resolution broadens each line by only a factor of 2--3. The spectral dispersion provides 5--6 pixels across each measured full width half maximum (FWHM), allowing centroid performance to be determined by signal-to-noise ratio (SNR) alone. The 1.55~\Rs\ field of view (FOV) is sufficient to sample different coronal conditions with a single slit position, while the 11--13 arcsec spatial resolution is sufficient to distinguish between different coronal features.  Near the limb where the corona is brightest, a 30 second exposure provides SNR sufficient to detect all of the lines and a 1 second exposure provides 5 km/s velocity resolution on \ion{Si}{10}.	
	
\revis{The instrument performance in Table \ref{tab:perf-req} was achieved by imposing requirements on the optical design, the precision of the optical alignment, the surface quality and efficiency of each optic, the image stability over each exposure, and the thermal design of the camera and spectrometer.  The optical design (Section \ref{sec:opt_sys}) provides the specified spectral range, spectral dispersion, spatial FOV, and spatial sampling. Achieving the spectral and spatial resolution requires root-mean-square (RMS) mirror surface figure errors below 1/20\textsuperscript{th} of a (633 nm) wave, alignment tolerances on the order of 20 \mic\ and 40 arcsec, (Section \ref{sec:align}), and an RMS image stability of 2 pixels or 4.6 arcsec (Section \ref{sec:imstab}). The SNR performance was obtained by purchasing mirrors and optical windows with high (85--98\%) optical efficiencies, cooling the spectrometer to 150 K (Section \ref{sec:bgreduction}), and purchasing an IR camera with high QE, low dark current and read noise, and a front end designed to minimize thermal emission (Sections \ref{sec:opt_sys} and \ref{sec:bgreduction}).}

\subsection{Design and Implementation}

\begin{figure*}[htp]
	\centering
	\includegraphics[angle=0,width=0.8\linewidth]{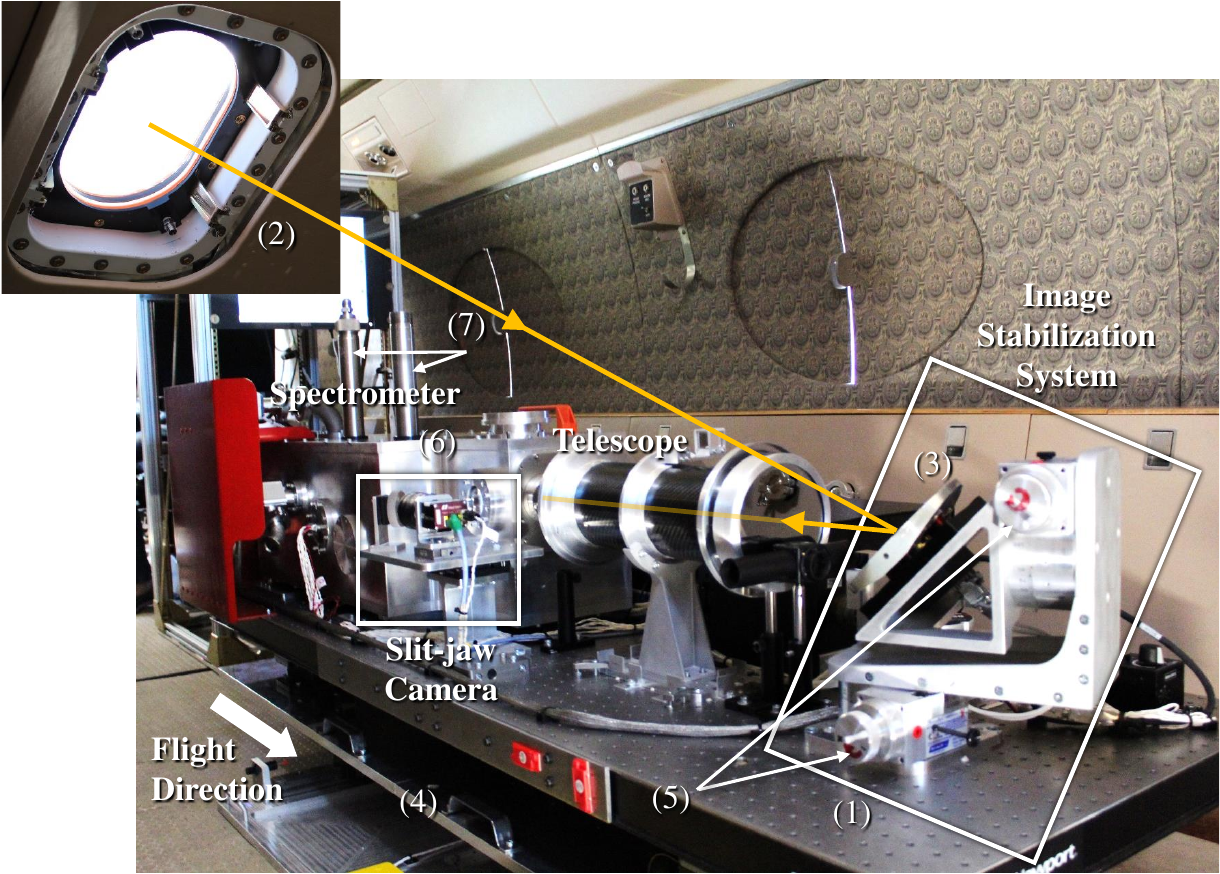}
	\caption{AIR-Spec on-board GV HIAPER. The instrument is installed on the floor \revis{on the port side of the cabin toward the front of the plane.} The image stabilization system, telescope, spectrometer, and slit-jaw camera are mounted on a vibration-isolated optical bench (1). Light enters through a small viewport \revis{(2)} on the upper \revis{starboard side} of the fuselage and is reflected into the telescope by a fast steering mirror (3).  The optical bench is translated fore-aft and left-right (4) and the mirror rotated about two axes (5) to facilitate alignment of the Sun, window, mirror, and telescope. The spectrometer optics are contained in a vacuum chamber (6) and cooled with liquid nitrogen (7).}
	\label{fig:instrument}
\end{figure*}

AIR-Spec consists of an image stabilization system, telescope, infrared spectrometer, and visible slit-jaw imager (Figure \ref{fig:instrument}).  The image stabilization system keeps the Sun fixed in the telescope field of view as the airplane moves. The telescope focuses onto a mirrored slit-jaw, where the light is divided into two channels.  Visible light reflected by the slit-jaw is imaged by a slit-jaw camera, while infrared light passing through the slit is dispersed by a diffraction grating and focused onto an infrared detector. \revis{The telescope entrance pupil is imaged onto the diffraction grating in a collimated beam.}

\revis{The telescope and spectrometer were designed in Zemax and implemented using a mixture of custom and modified COTS components, as detailed in Section \ref{sec:opt_sys}.  Alignment sensitivities for each individual optic were determined using Zemax and an error budget was compiled in order to define the requirements on alignment, optical surface quality, and image stability. In lieu of a full stray light analysis (e.g. using non-sequential Zemax or FRED), the cryostat design (Section \ref{sec:bgreduction}) was informed by hand calculations that estimated the photons scattered and emitted onto the detector from various sources inside the spectrometer.}

\subsubsection{Optical System}
\label{sec:opt_sys}
\begin{figure*}[htp]
	\centering
	\includegraphics[angle=0,width=1\textwidth]{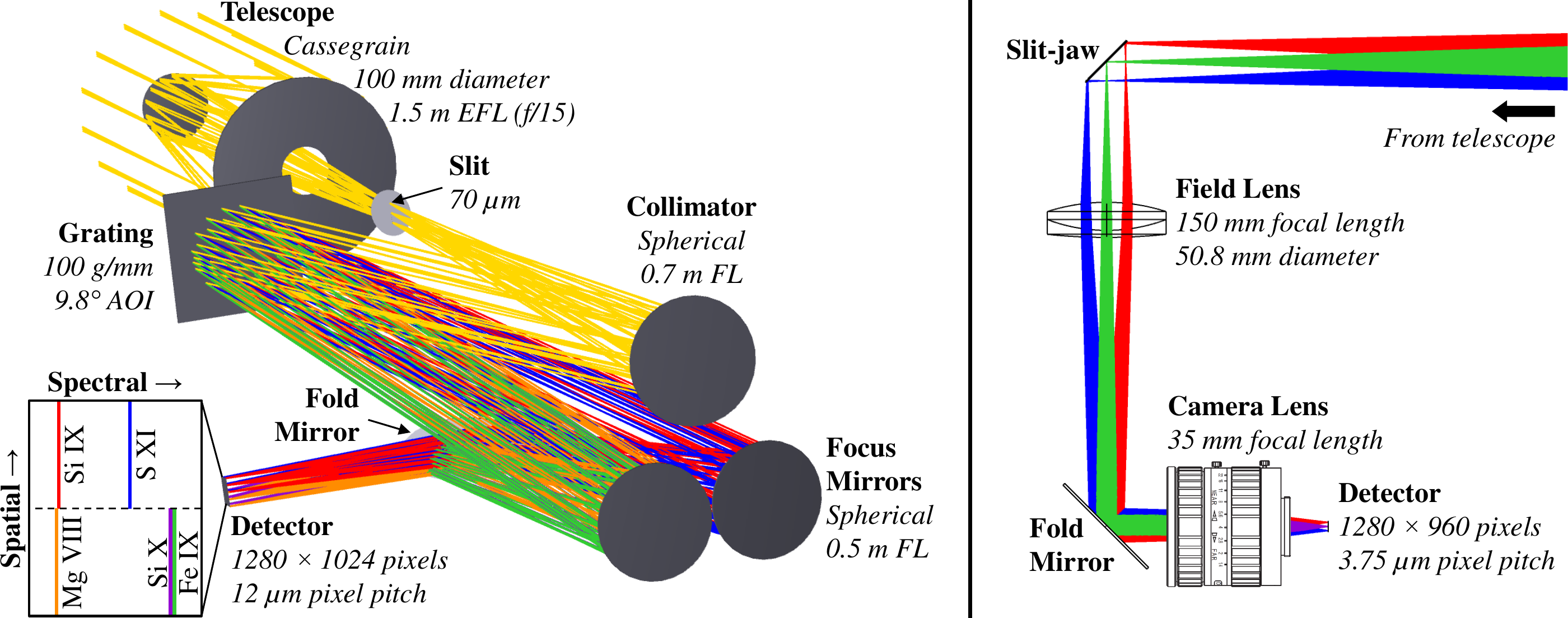}
	\caption{Left: Raytrace of the AIR-Spec telescope and spectrometer, including the layout of emission lines on the infrared detector. \revis{Different ray colors indicate different wavelengths}. Right: Raytrace of the slit-jaw camera optics. \revis{Different ray colors indicate different field angles.}}
	\label{fig:raytrace}
\end{figure*}

Figure \ref{fig:raytrace} shows a raytrace of the AIR-Spec telescope and spectrometer.  Light is collected by the f/15 Cassegrain telescope (10 cm primary, 1.5 m focal length) and focused onto a 70 \mic\ \revis{(9.6 arcsec)} entrance slit. \revis{The slit is cut into an aluminum coating on a 2 mm thick sapphire substrate.  The aluminum coating is on the front surface of the substrate and reflects light into a slit-jaw camera. Stray reflections from the uncoated back surface introduced spectral artifacts in the 2017 data that were misinterpreted as emission lines \citep{Samra2019}. A backside metal shield was later added to absorb these reflections, as described in Section \ref{sec:lessons}.}

The light exiting the slit is collimated by a 0.7 m focal length spherical mirror, and the collimated light is incident on a reflective planar diffraction grating with 10 \mic\ groove spacing. \revis{The grating is a replica on a Pyrex substrate that was custom-made by Thorlabs. It has 10 \mic\ groove spacing, a blaze angle of 9.7$^\circ$, an aluminum coating, and an efficiency that ranges from 23\% to 79\% at the five wavelengths of interest.} The grating is used in a near-Littrow configuration, so that the diffracted wavelengths are nearly coincident with the incoming light.   The five wavelengths of interest are diffracted in two groups, second order 1.92~\mic\ rays near first order 3.93 \mic\ rays, and second order 1.43 \mic\ rays near first order 2.86 \mic\ and 3.03 \mic\ rays.   Two 0.5 m focal length spherical mirrors focus the two wavelength groups onto the top and bottom of the detector. The 3~\mic\ and 4~\mic\ channels include second order light at 1.5~\mic\ and 2~\mic, respectively. The detector layout is shown in Figure \ref{fig:raytrace}. The spectral and spatial characteristics of each channel are listed in Table \ref{tab:perf-req}. 

\revis{Custom optics were purchased for the telescope primary and secondary, the spectrometer collimator and focus mirrors, and the diffraction grating.  The spectrometer fold mirror is an off-the-shelf part that was cut to size after purchase, and the slit-jaw substrate is an off-the-shelf sapphire window with a custom coating.}

\revis{The infrared camera is an IRC912 from IRCameras (\url{https://ircameras.com}) with a custom focal plane (Lockheed Martin Santa Barbara Focalplane model SBF204), multiband cold filter, cold stop, and antireflection coated dewar window. 	This was the only camera within our budget that met our requirements for sensitivity, noise, resolution, and format. Its specifications are summarized in Table \ref{tab:ircamera}.}

\revis{The SBF204 readout integrated circuit, with 50,000 electron well depth and $<$32 electron read noise, is specifically designed for low photon flux applications. The average quantum efficiency of the InSb sensor is $>$90\% due to its wideband 1--5.3 \mic\ antireflection coating. The cold filter and cold stop, both operating at 59~K, shield the focal plane from thermal photons outside the field of view and/or passband. The focal plane itself also runs at 59 K, reducing the dark current to about 65,000 e$^-$/s/pixel. }
	
The focal plane has 1024 pixels along the spectral dimension and 1280 pixels along the spatial dimension (640 pixels per channel).  The 12 \mic\ pixel pitch provides a linear dispersion of about 2.4 \AA/pixel in first order (1.2 \AA/pixel in second order) and a plate scale of 2.3 arcsec/pixel.  The spectral width of each channel is about 2400 \AA\ in first order and 1200 \AA\ in second order.  The slit length is $0.4^{\circ}$ (1.5 \Rs).  During the eclipse, the IR camera operated at a cadence of 15 frames/sec.  Most of the coronal data were collected with a 60 ms exposure time.
	
\begin{table}[!htp]
	\raggedright
    \caption{\revis{Characteristics of the infrared camera.}}
    \begin{tabular}{p{0.39\columnwidth}p{0.30\columnwidth}}
        \tableline
        \revisb{Material} & \revisb{InSb} \\
        \revisb{Wavelength range} & \revisb{1.0--5.3 \mic} \\
       	\revisb{Operating temperature} & \revisb{59 K} \\
    	\revisb{Format} & \revisb{$1280 \times 1024$ pixels}\\
    	\revisb{Pixel pitch} & \revisb{12 \mic}\\
	    \revisb{Well depth} & \revisb{50,000 e$^-$}\\
    	\revisb{Quantum efficiency} & \revisb{$>90$\%}\\
    	\revisb{ADC gain} & \revisb{3.25 e$^-$/DN}\\
    	\revisb{Readout noise} & \revisb{$<32$ e$^-$}\\
    	\revisb{Dark current} & \revisb{65,000 e$^-$/s/pixel}\\
	\tableline
	\end{tabular}
	\label{tab:ircamera}
\end{table}

\begin{figure*}[htp]
	\centering
	\includegraphics[angle=0,width=1\textwidth]{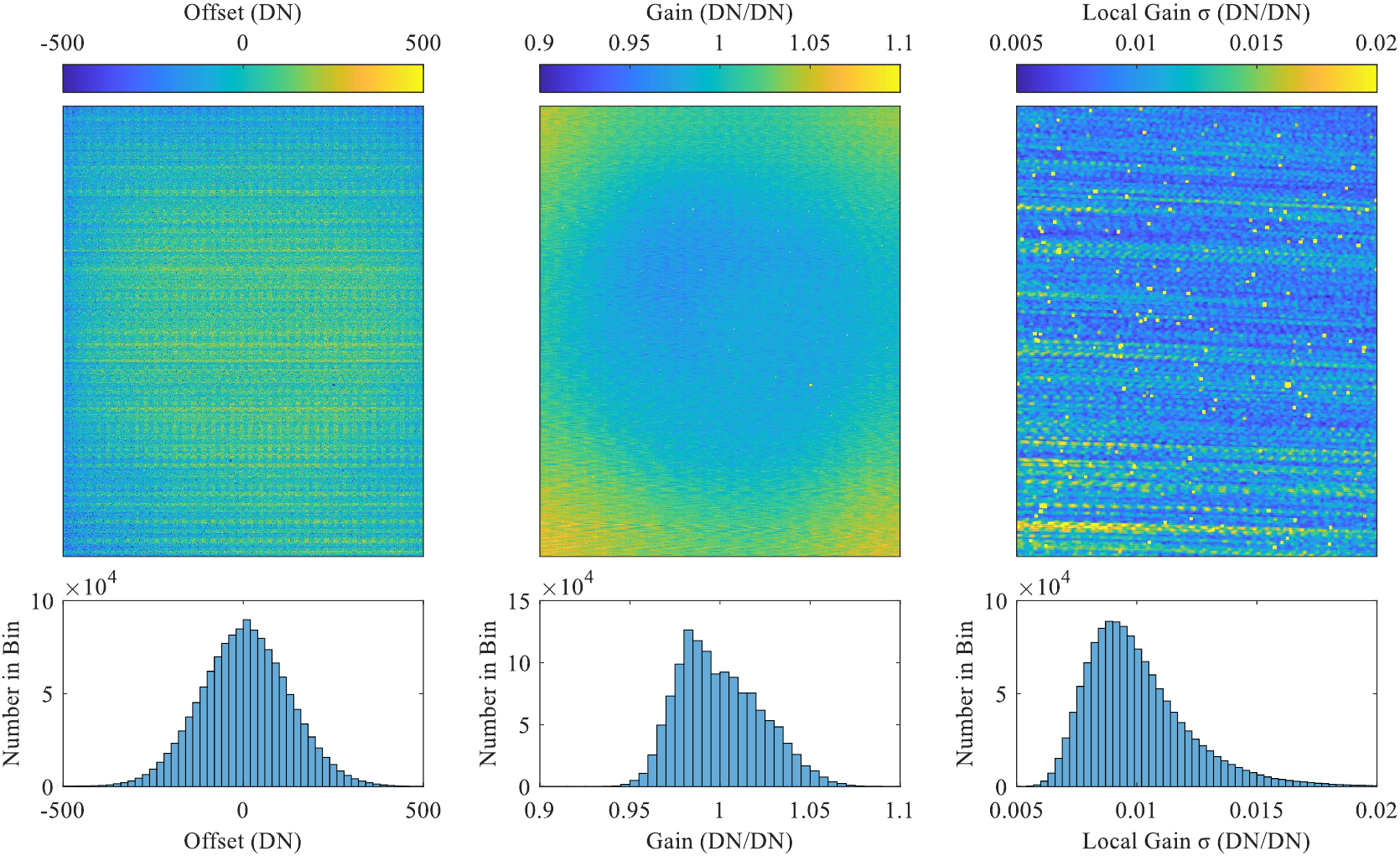}
	\caption{\revis{Maps and histograms of spatial non-uniformity across the detector. The left and middle panels show the offset and gain needed to perform a two-point non-uniformity correction at every pixel. The right panel shows the gain standard deviation in each local 9$\times$9 pixel neighborhood.}}
	\label{fig:nuc}
\end{figure*}

\revis {The IR detector has high spatial uniformity, as judged from flat field data taken by the manufacturer at 2~ms exposure time.  (It was not possible to introduce a Lambertian source into the cold cryostat in order to repeat this test at a more representative exposure time.)  The first two panels in Figure \ref{fig:nuc} show the spatial variation of the two-point non-uniformity correction  across the detector.  98\% of pixels have an offset between -300 and 300 DN and a gain between 0.95 and 1.05 DN/DN.  Local variations in the gain are much smaller, as seen in the third panel.  The gain standard deviation is less than 0.02 DN/DN in 99\% of 9$\times$9 pixel neighborhoods.}

The white light slit-jaw camera provides context imagery for the infrared spectra. Light from the mirrored slit-jaw is re-imaged by a monochrome visible camera, in which the slit appears as a dark line superimposed on the corona.  Both the white light camera and its lens are commercially available products. \revis{They were chosen to match or exceed the field of view, spatial resolution, and cadence of the spectrometer and to make software integration as simple as possible.} The camera is a Prosilica GC1290 from Allied Vision Technologies, with $1280\times960$ 3.75 \mic\ pixels. \revis{It has a gigabit ethernet interface, a global shutter, a maximum frame rate of 33 Hz, a quantum efficiency of 30--60\% at visible wavelengths, and a 12 bit ADC.} The f/1.4, 35 mm camera lens is a Fujinon CF35HA-1. A field lens ensures that the off-axis rays from the slit-jaw are not vignetted at the camera lens.  The field lens is a 150 mm achromat that serves as the vacuum chamber exit window for the light reflected from the slit-jaw.  The \revis{lens-camera system} has a plate scale of 2.3 arcsec/pixel (equivalent to the IR camera plate scale), a 2.3 \Rs\ field of view (1.5 times larger than the IR camera FOV), and a cadence of 33 frames/sec (about twice the IR camera cadence).  During totality, the exposure time was set to 1 ms to provide a well-exposed inner corona. A ray trace of the slit-jaw camera optics is shown in Figure \ref{fig:raytrace}.

The two main challenges of implementing the AIR-Spec optical design were (1) pointing the telescope at the Sun stably and continuously and (2) minimizing the level of the dark instrument background. Pointing and stabilization were achieved by actively controlling the line of sight (LOS) with a fast steering mirror and manually adjusting the mirror and table to compensate for changes in viewing geometry. The instrument background was reduced by cooling the spectrometer optics and infrared camera. Figure \ref{fig:instrument} shows AIR-Spec installed in the GV cabin with image stabilization and cooling system components identified.

\subsubsection{Pointing and Image Stabilization} \label{sec:imstab}
During the 2017 eclipse, AIR-Spec observed through a $150\times220$ mm double-paned sapphire viewport on the \revis{upper starboard} side of the aircraft cabin. Sapphire was chosen because it provided high transmission (around 85\% per windowpane) from visible to mid-IR wavelengths. \revis{To minimize its birefringence, the sapphire was cut with its fast axis normal to the plane of the window. For cost and schedule reasons, the sapphire panes were designed to conform to a standard GV viewport that was already certified to fly.  This required that all four semi-reflective surfaces be parallel, resulting in an etalon effect that produced fringes in one of the spectrometer channels (see Section \ref{sec:fringe})}.

The instrument was mounted to the floor on the \revis{port} side of the cabin, where a fast-steering mirror on the optical bench directed sunlight from the window into the telescope (Figure \ref{fig:instrument}). Since both the field of view through the window and the mirror range of travel were limited, the instrument was positioned fore--aft and left--right using linear sliders and the mirror normal was aligned using two manual rotary adjusters on its mount.  A custom alignment tool similar to a telescope reflex sight was used to place the Sun in the telescope through the window as the instrument was translated and the mirror was rotated. The alignment tool superimposed an LED referencing the telescope LOS onto a view of the Sun through the aircraft window. 

\begin{figure*}[htp]
	\centering
	\includegraphics[angle=0,width=1\textwidth]{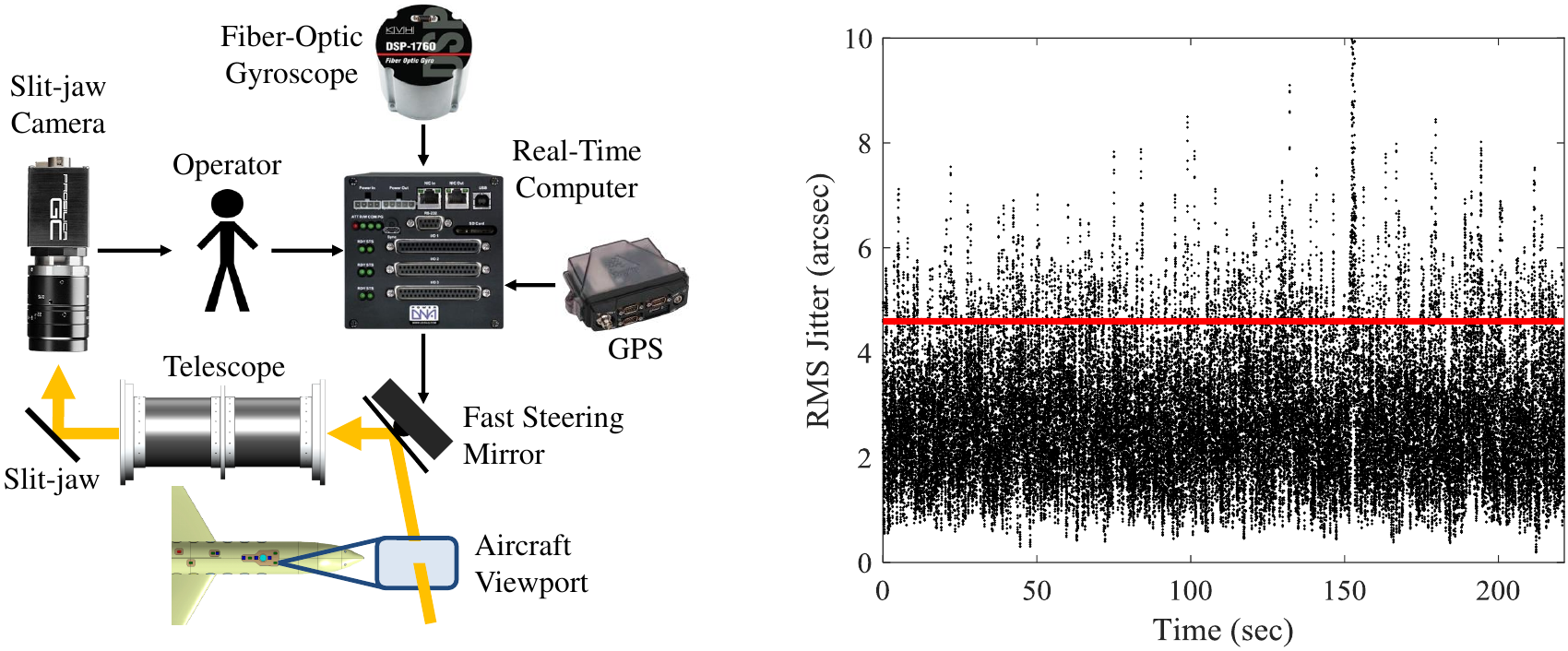}
	\caption{Image stabilization system block diagram (left) and performance (right). RMS jitter for each 60 ms exposure is plotted in black. The 2 pixel Nyquist limit is the red line.  92\% of exposures have jitter below the Nyquist limit.}
	\label{fig:img-stab}
\end{figure*}

Image stability was addressed in two ways.  First, the optical bench was isolated from the effects of high-frequency airframe vibration by six tuned isolators placed between it and the aircraft floor.  Second, the remaining low-frequency perturbations were compensated by a closed-loop fast steering mirror that fed a stabilized beam into the telescope. The mirror was dynamically positioned so that the line of sight from the Sun was normally incident on the telescope.  The image stabilization requirement was 4.6 arcsec (2 pixels) RMS over each 60~ms camera exposure. During the four minute eclipse observation, 92\% of exposures achieved this requirement (Figure \ref{fig:img-stab}). 

The image stabilization is shown schematically in Figure \ref{fig:img-stab}. Computations are performed at 500 Hz by a real-time computer. The mirror command is calculated from three inputs: the eclipse ephemeris (computed using GPS location and time), aircraft attitude given by the integrated gyroscope rates, and slit position coordinates from the operator.  In order to position the slit, the operator monitors a video of the white light corona from the slit-jaw camera. 

\subsubsection{Thermal Background Reduction}\label{sec:bgreduction}

Because of their close proximity to the focal plane, the slit-jaw, collimator, grating, focus mirrors, and camera fold mirror emit enough radiation at room temperature to overwhelm the coronal signatures. Radiation from the slit-jaw is especially significant because it is imaged onto the focal plane by design.  In order to maximize signal-to-noise ratio, the slit-jaw and all subsequent optics are cooled to 150 K. To achieve this while keeping the optics dry and stable, the entire spectrometer is housed in a vacuum chamber at a pressure below $10^{-3}$ Torr. \revis{The chamber floor is the top of a liquid nitrogen dewar, which serves as the heat sink for the components inside the chamber. The 2 L dewar has a hold time of over one hour once the system reaches thermal equilibrium.}

\begin{figure*}[htp]
	\centering
	\includegraphics[angle=0,width=1\textwidth]{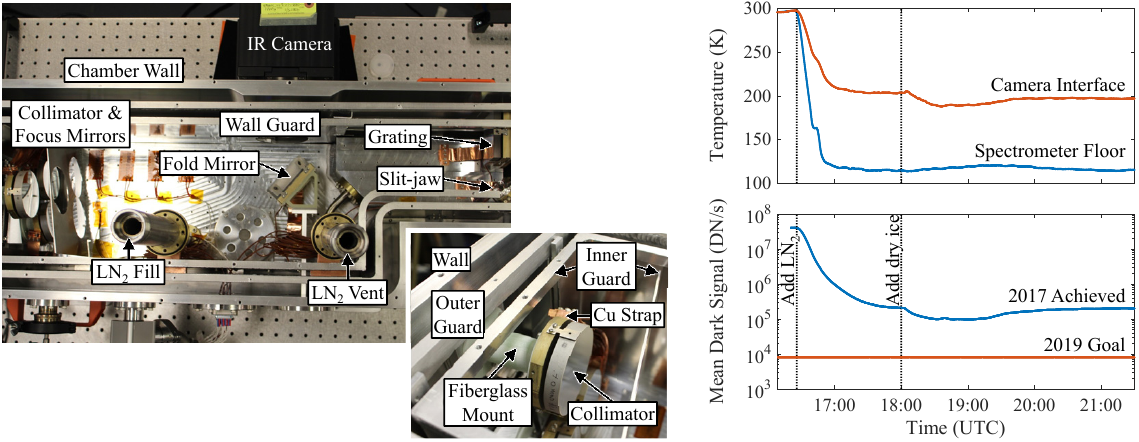}
	\caption{Thermal background reduction. Left: Top-down view of the AIR-Spec vacuum chamber before it was blackened to reduce stray light. The spectrometer optics and IR camera are labeled. Liquid nitrogen enters through the LN$_2$ fill port and nitrogen gas is released through the vent. Copper straps connect the optics to the floor, which is the top of the liquid nitrogen dewar. Inset: Detail of the collimator mount and cooling strap.  To maximize stability, each optic is mounted to the warm chamber wall with G10, a thermally isolating fiberglass composite. A thermally conducting copper strap attaches the optic to the cold spectrometer floor. Inner and outer wall guards block radiation from the warm chamber wall. Right: Effect of cooling the instrument.}
	\label{fig:thermal}
\end{figure*}

  Inside the vacuum chamber, the optics are mounted to thermally isolated \revis{tungsten temperature-spreader} plates and attached with copper straps to the floor. \revis{Tungsten was chosen for its combination of high thermal conductance (164 W/m/K) and low coefficient of thermal expansion (CTE, 4.4 ppm/$^\circ$C).} Wall guards \revis{chilled to $<$150 K} shield the detector from thermal radiation from the warm walls, and the wall guards and floor are blackened to reduce reflectivity. Figure \ref{fig:thermal} shows the inside of the vacuum chamber, including the spectrometer optics, liquid nitrogen ports, and chilled wall guards. Two feedthrough micrometers on the focus mirrors (not shown) allow cold alignment of the two channels on the detector. The inset shows in more detail how the optics are mounted and cooled.

The infrared camera includes several features to help minimize its dark background. \revis{As described in Section \ref{sec:opt_sys}}, the focal plane was chilled \revis{to 59 K} by a closed-cycle cooler \revis{in order to reduce the dark current to 65,000 e$^-$/s/pixel}. The thermal background was reduced by a cold \revis{stop}, which limited the field of view, and a cold bandpass filter, which removed light outside the instrument passbands. 

\revis{In spite of these features, we observed significant thermal emission originating from the front end of the camera. This was a much larger contributor to the dark background than either dark current or thermal emission from inside the spectrometer. We created a thermal model of the camera housing and the nearby wall of the spectrometer, compared the modeled results to thermocouple measurements, and determined that heat from the camera cryocooler was conducting into the front of the camera housing and emitting onto the detector.  We reduced thermal emission from the housing in two ways: we packed the camera-spectrometer inteface in dry ice, and we placed a cold baffle assembly in contact with the camera entrance window.  The aluminum/copper baffle attaches directly to the cold inner wall guard and uses thermally conductive gasketing material to make a thermal connection to the camera window.}

\revis{Chilling the spectrometer with liquid nitrogen and the camera housing with dry ice reduced the thermal background by a factor of 400 compared to room temperature operation (Figure \ref{fig:thermal}). During the 2017 eclipse,} the background level was on the order of $10^5$ DN/sec, and a 60 ms exposure time limited the background to half the 15,000~DN well depth.  \revis{Ahead of the 2019 eclipse, IRCameras completed a redesign and rebuild of the camera that reduced the background by an additional factor of 30. Section \ref{sec:lessons} describes the camera modifications needed to achieve this improvement.}

\subsection{Optical Alignment} \label{sec:align}
The AIR-Spec alignment took place in five steps.  First, the telescope secondary was aligned to the primary.  The internal spectrometer optics were aligned next, and then the slit-jaw camera was aligned to the slit-jaw.  Finally, the telescope was aligned and focused relative to the spectrometer, and the image stabilization components were aligned to the telescope.

The purpose of the optical alignment was to minimize the instrument point-spread function and center and focus the image on each camera. Because small misalignments could be compensated by adjusting focus, it was sufficient to align each optic to several hundred microns in centration and several arcminutes in tilt.  With the system aligned to this level, misalignment had a negligible effect on point-spread function (PSF) compared to imperfections in the mirror surfaces.

\subsubsection{Telescope Alignment}
The method for aligning the AIR-Spec telescope was modeled on the alignment procedure for the Atmospheric Imaging Assembly (AIA) telescopes \revis{(\citealp{Lemen2012}; W. Podgorski private communication to V. Marquez 2017)}. Before the primary and secondary mirrors were aligned, a corner cube was positioned to reference the mechanical boresight of the telescope tube. Two cross-hair reticles were mounted at either end of the tube, one in the primary mirror hole and the other in place of the secondary. An alignment telescope was used to find the line through both reticles, and this line was defined as the mechanical boresight.  With the alignment telescope in auto-collimation mode, a retro-reflecting flat mirror was aligned to both telescopes. Finally, the corner cube was aligned to the retro flat using a theodolite and bonded into place.

\begin{figure*}[htp]
	\centering
	\includegraphics[width=1\textwidth]{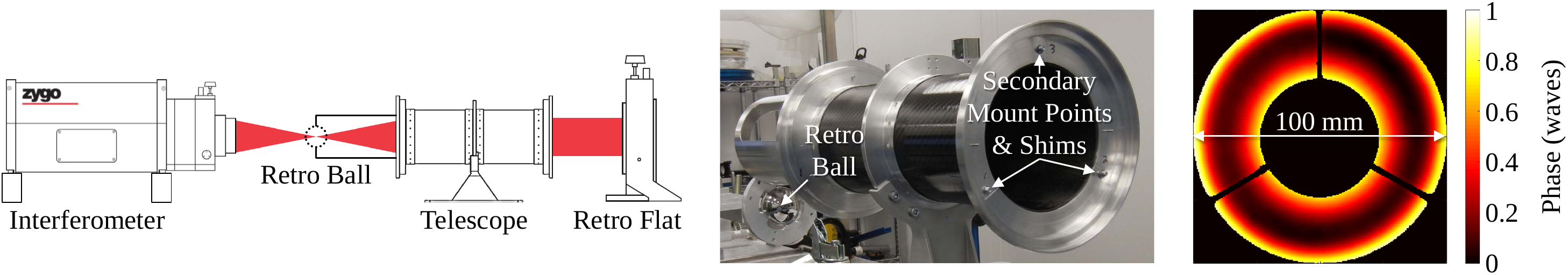}
	\caption{Telescope alignment. Left: Optical setup, including the Zygo interferometer, retro ball, AIR-Spec telescope, and retro flat \citep{Zygo}. Center: Aligning the secondary mirror. The retro ball is off-axis, allowing the Zygo laser beam to pass through the system. The secondary is tilted and focused by shimming with washers at the three points indicated. Right: Double-pass wavefront for the final telescope alignment, after zeroing tip and tilt.  The phase variations are due to a combination of defocus and the $\lambda$/20 RMS surfaces of the primary and secondary.}
	\label{fig:tel-align}
\end{figure*}

The secondary mirror was aligned to the primary with the telescope in the double-pass configuration shown in Figure \ref{fig:tel-align} (left), from a Zygo application note on typical interferometer setups \citep{Zygo}.  A Zygo interferometer (model GPI-4-XP 512, 633 nm wavelength) fed the telescope and precisely measured the tip, tilt, and defocus of the returned wavefront, which was reflected back through the system by the retro flat. The focal point was defined by the center of a removable retro-reflecting ball mounted to the back of the telescope.

The interferometer was outfitted with an f/7.2 Dynaflect transmission sphere, suitable for aligning reflective surfaces, and the focus of the spherical beam was placed at the nominal telescope focus by translating the interferometer until the wavefront was approximately flat. The final wavefront had a peak-to-valley height of 0.76 waves and a power term of 0.13 waves, corresponding to about 3 \mic\ lateral displacement and 30 \mic\ axial displacement of the focal point from the center of the retro ball. 

Next, the retro ball assembly was removed from the rear telescope flange and mounted upside down, maintaining the tilt of the telescope while allowing the interferometer beam to pass through it (Figure \ref{fig:tel-align}, center).  The secondary mirror was shimmed with washers at its three mounting points, resulting in a final tilt of 28 waves (270 \mic\ lateral displacement) and defocus of 0.23 waves (260 \mic\ axial displacement).  The right plot in Figure \ref{fig:tel-align} shows the double-pass wavefront after zeroing tip and tilt by adjusting the retro flat. In addition to defocus, the variations across the wavefront come from imperfections in the mirror surfaces.

\subsubsection{Spectrometer Alignment} \label{sec:alignspec}
The spectrometer alignment was complicated by the number of optical elements and the non-axial light path.  However, the long depth of focus (0.5 mm) and the use of spherical mirrors provided relatively loose alignment tolerances that were achieved using shims and measurement tools such as a shear plate, a ruler, and the IR camera. Because all of the spectrometer optics were either planar or spherical, decenter could be compensated by tilt with little effect on PSF. Therefore, it was sufficient to center the light path on each optic to about 0.5 mm and precisely align the image on the detector by tilting the focus mirrors with a set of micrometers. 

\begin{figure*}[htp]
	\centering
	\includegraphics[width=0.9\textwidth]{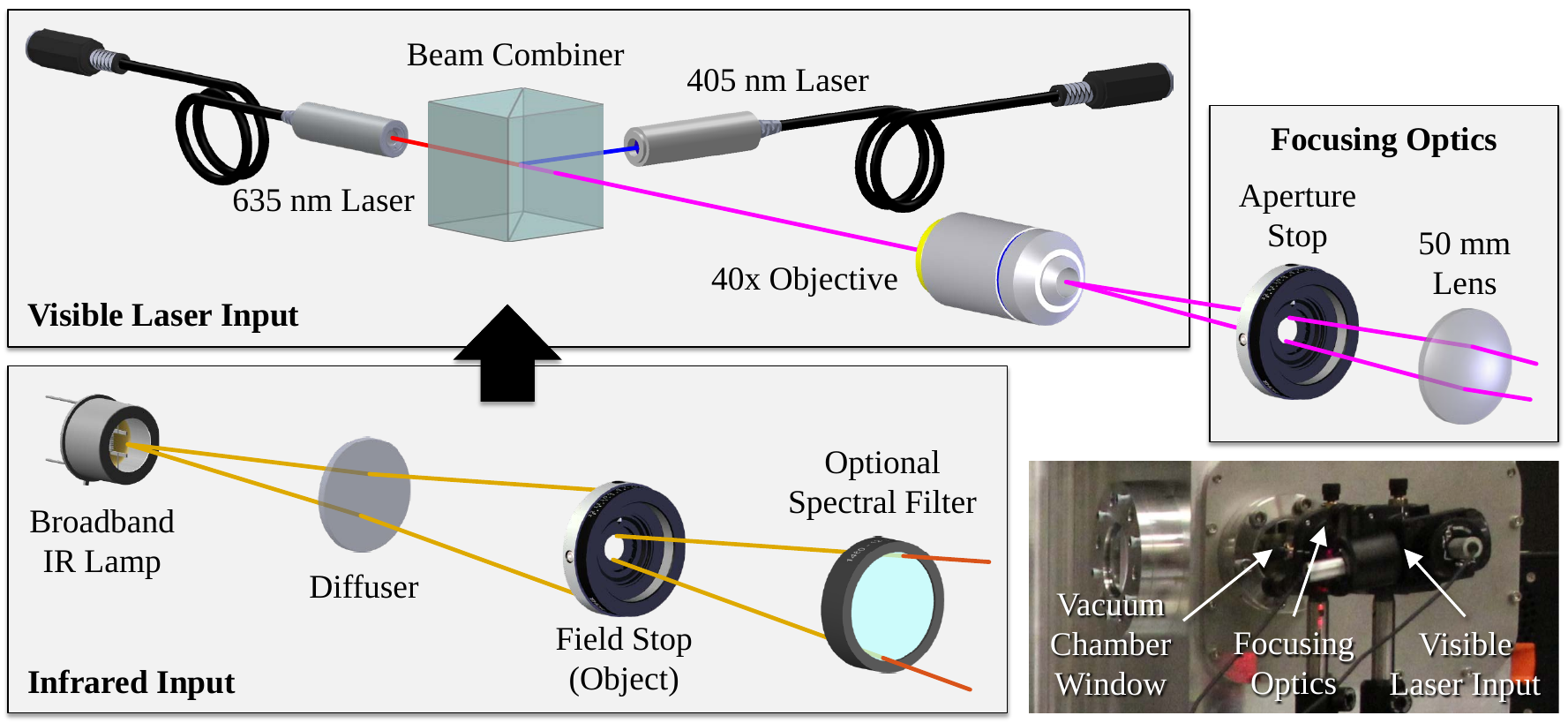}
	\caption{Input options for spectrometer alignment. The focusing optics are fed by either two visible lasers or a broadband IR lamp.  The focusing optics image the object (microscope objective spot or IR-illuminated field stop) onto the slit through the vacuum chamber window (lower right).}
	\label{fig:spec-input}
\end{figure*}

Two input assemblies were constructed to allow alignment in both visible and infrared light (Figure \ref{fig:spec-input}).  The visible laser assembly combined red (635 nm) and blue (405 nm) lasers with a 50-50 beamsplitter. A 40x microscope objective was used to create a fast divergent beam. The infrared assembly consisted of a field stop (iris aperture) illuminated by a diffuse broadband IR source.  One of several exchangeable filters could be included to tailor the spectral output. A 50 mm MgF$_2$ lens focused light from the operational assembly, imaging the object (microscope objective spot or IR-illuminated field stop) onto the spectrometer slit-jaw.  The lens was placed approximately 75 mm from the object and 150 mm from the slit, providing a magnification of about 2. The f number was set by another iris, which acted as an aperture stop. The spectrometer was aligned according to the following procedure:
\begin{enumerate}\itemsep0pt
	\item  A 50 \mic\ pinhole was mounted in place of the slit-jaw. The laser input assembly (with the blue laser off) was aligned through the pinhole to the center of the collimating mirror, defining the optic axis.
	
	\item \textit{Pinhole focus:} A shear plate was used to check the beam from the collimating mirror. The mirror was shimmed until the shear plate fringes were parallel to the fiducial line, indicating that the pinhole was coincident with the mirror focal point (Figure \ref{fig:spec-align}a).
	
	\begin{figure*}[htp]
		\centering
		\includegraphics[angle=0,width=1\textwidth]{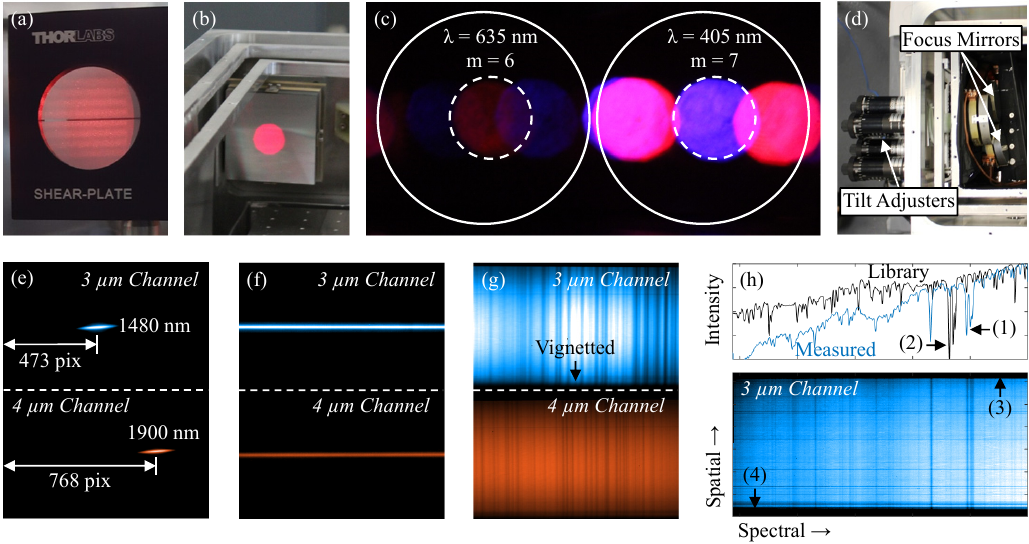}
		\caption{Alignment of the spectrometer optics. (a) Pinhole focus. The shear plate fringes are parallel to the fiducial line, indicating that the pinhole is at the focus of the collimator. (b) Collimator tilt. The collimator has been tilted to center the collimated beam on the grating. (c) Grating tilt. The grating has been tilted to properly align the red and blue dispersed beams on the focus mirrors. (d) Feedthrough micrometers for tilting the focus mirrors. (e) Focus mirror tilt. Each focus mirror has been tilted to center the pinhole image vertically and properly align the passbands horizontally. (f) Grating rotation. The grating has been clocked so that the broadband image of the pinhole is approximately horizontal in both channels.  (g) Slit rotation. The slit has been clocked so that the atmospheric absorption features are approximately vertical in both channels.  (h) Cold alignment. Screenshot from the IR  visualization software during in-flight alignment of the focus mirrors. The lower plot shows the 3 \mic\ spectral~$\times$~spatial image and the upper a spectral cross-section of the image.  Spectral alignment is achieved when photospheric absorption features in the measured spectrum (1) overlap with those in the library spectrum (2). Spatial alignment is achieved when light from the slit is centered vertically in the detector, resulting in a dark band at the top (3) and bottom (4).}
		\label{fig:spec-align}
	\end{figure*}
	
	\item \textit{Collimator tilt:} The 50 \mic\ pinhole was replaced by a 400 \mic\ pinhole and the blue laser was turned on. The collimating mirror was tilted about two axes until the beam was centered on the diffraction grating (Figure \ref{fig:spec-align}b).
	
	\item \textit{Grating tilt:} The grating was tilted about two axes until the 6\textsuperscript{th} order beam from the red laser ($m\lambda=3.810$~\mic) and the 7\textsuperscript{th} order beam from the blue laser ($m\lambda=2.835$~\mic) landed in their intended locations on the two focus mirrors (Figure \ref{fig:spec-align}c).
	
	\item The laser assembly was replaced by the IR assembly. The field stop was adjusted to fully illuminate the pinhole while underfilling the 25 mm pinhole substrate, to prevent stray light from reaching the detector. The illuminated pinhole produced a bright line in each channel of the IR camera. 
	
	\item \textit{Focus mirror tilt:} Narrowband filters centered at 1480 nm and 1900 nm were added one at a time.  Using the feedthrough micrometers (Figure \ref{fig:spec-align}d), each focus mirror was tilted about two axes until the line was centered vertically in the channel and the spectral passband was centered at the intended horizontal pixel (Figure \ref{fig:spec-align}e). 
	
	\item \textit{Grating rotation:} The narrowband filters were removed. The grating was rotated in-plane until the line in each channel was approximately horizontal (Figure \ref{fig:spec-align}f).
	
	\item \textit{Camera focus:} The 400 \mic\ pinhole was replaced by the 50 \mic\ pinhole (an approximate point source). The focus of the pinhole image was checked in both channels. The point-spread function was about as wide as predicted, so the focus mirrors were not moved. 
	
	\item \textit{Slit rotation:} The 400 \mic\ pinhole was removed and the slit-jaw was mounted in its place. The field stop was adjusted to overfill the slit while underfilling the slit-jaw.  The slit-jaw was rotated in-plane until the observed atmospheric absorption features were approximately vertical in both channels (Figure \ref{fig:spec-align}g). Although the alignment was performed indoors, the absorption was measurable due to the long path length in the open vacuum chamber. 
	
	\item \textit{Slit focus:} The top and bottom edges of the slit image were checked in the IR camera.  The top edge was well-focused, indicating that the slit-jaw did not need to be adjusted.  Light from the bottom of the slit was vignetted by the slit-jaw mount, reducing the slit length by 10\% (Figure \ref{fig:spec-align}g).  This issue \revis{was} corrected before the 2019 eclipse.
	
	\item \textit{Cold alignment:} The spectrometer was sealed, pumped down, and chilled.  As designed, the image focus, spatial rotation, and spectral rotation remained unchanged. The image shifted spectrally and spatially, and this was corrected by tilting the focus mirrors. The final alignment of the focus mirrors was performed in flight, using absorption lines in the solar photosphere to place the spectrum on the detector (Figure \ref{fig:spec-align}h).
\end{enumerate}

\subsubsection{Slit-jaw Camera Alignment} \label{sec:alignsjcamera}
\begin{figure}[htp]
	\centering
	\includegraphics[width=0.9\linewidth]{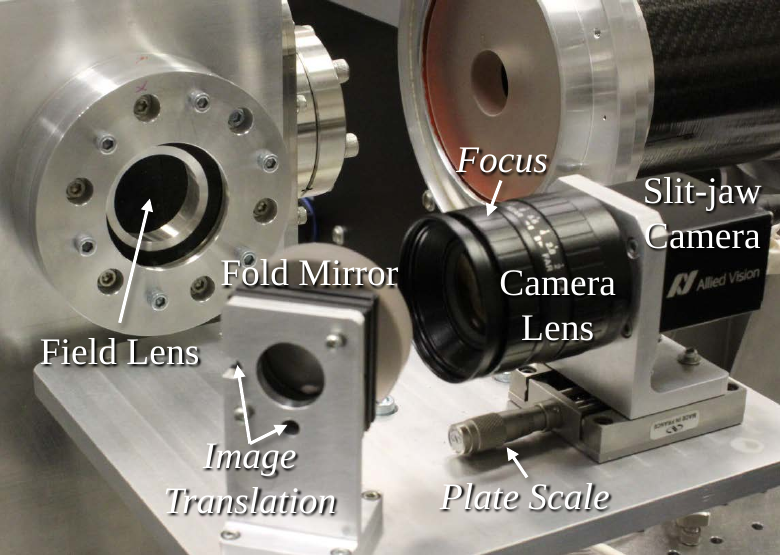}
	\caption{Slit-jaw camera and associated optics.}
	\label{fig:sj-optics}
\end{figure}

Once the position of the slit-jaw was finalized, the slit-jaw camera was aligned to it.  Figure \ref{fig:sj-optics} shows the slit-jaw camera and associated optics, with the available adjustments labeled. The slit-jaw itself is out of sight inside the vacuum chamber. The fold mirror was tilted about two axes to center the slit in the image.   The plate scale was set by translating the camera and lens as a unit, and the image was focused using the focus adjustment on the camera lens. The target plate scale was 2.31 arcsec/pixel, equivalent to that of the IR camera. Due to limited travel in the translation stage, the maximum available plate scale was 2.26 arcsec/pixel (Section \ref{sec:platescale}).

\subsubsection{Telescope and Spectrometer Alignment}
\begin{figure*}[htp]
	\centering
	\includegraphics[width=0.8\textwidth]{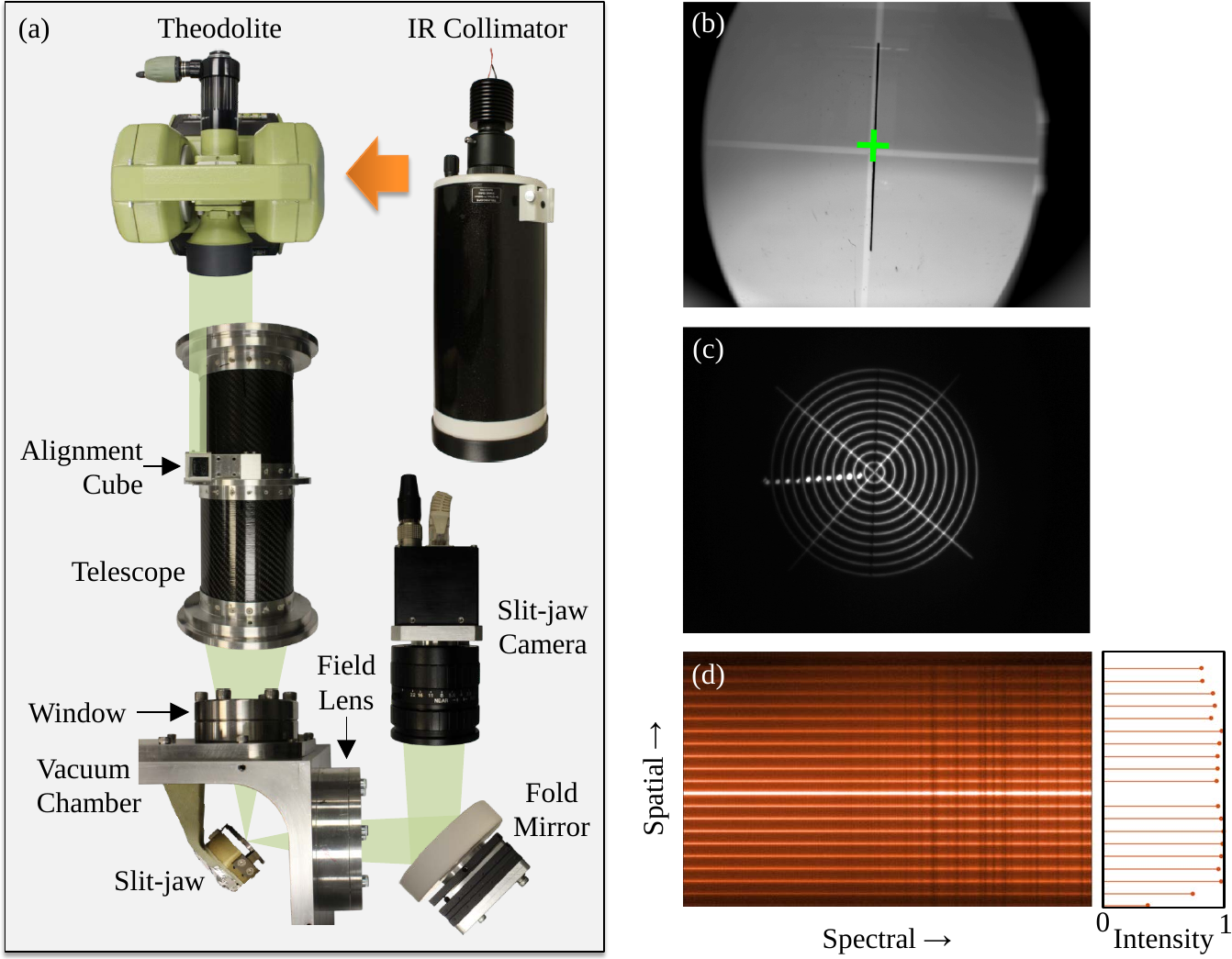}
	\caption{Aligning the telescope to the spectrometer. (a) Diagram of components. The theodolite was used to align tip, tilt, and focus, while the IR collimator was used to check centration. (b) Slit-jaw image of the theodolite crosshair. The green plus sign marks the center of the slit, 40 arcsec from center of the theodolite crosshair. The slit and theodolite crosshair are both in focus. (c) Slit-jaw image of a concentric circle target from the IR collimator. (d) 4 \mic\ IR image of the concentric circle target. The intensity is uniform over the length of the slit, confirming that the telescope pupil is centered on the spectrometer mirrors.}
	\label{fig:tel-spec-align} 
\end{figure*}

The telescope was aligned to the spectrometer in tip, tilt, focus, and horizontal and vertical translation.  Alignment took place with the spectrometer evacuated, cold, and internally aligned, ensuring that all optical elements were in their flight positions.  Tip, tilt, and focus were aligned using a theodolite, and translation was checked using an infrared collimator. Figure \ref{fig:tel-spec-align} shows the alignment scheme and results.

The telescope was placed in its nominal location in front of the spectrometer and fed with a collimated beam from the theodolite.  A piece of the theodolite beam was retro-reflected by the telescope alignment cube, allowing the theodolite to be precisely aligned to the telescope line of sight. The theodolite output was focused onto the slit-jaw by the telescope and re-imaged by the slit-jaw camera (Figure \ref{fig:tel-spec-align}a).

The telescope was adjusted in tip, tilt, and focus until the image of the theodolite crosshair was centered and focused on the slit (Figure \ref{fig:tel-spec-align}b).  Each time the telescope moved, the theodolite had to be realigned to the telescope alignment cube. This iterative process continued until the center of the theodolite crosshair was less than one arcminute from the center of the slit. In Figure \ref{fig:tel-spec-align}b, the green plus sign marks the center of the slit.  The center of the theodlite crosshair is displaced by 40 arcsec.

Next, the infrared image was checked for vignetting resulting from decenter of the telescope pupil on the spectrometer mirrors.  Due to diffraction at the slit, the telescope overfilled the spectrometer mirrors horizontally but not vertically. As the vertical direction was more sensitive to decenter, centration was assessed by checking the intensity of the IR image along the slit. The theodolite was replaced with an infrared collimator (Figure \ref{fig:tel-spec-align}a), which fed the telescope with collimated light from a concentric circle target illuminated by a broadband IR lamp. Figure \ref{fig:tel-spec-align}c shows the slit-jaw image of the target with the slit through the center.  Figure \ref{fig:tel-spec-align}d shows the 4 \mic\ infrared image and the spectrally integrated intensity at each bright circle sampled by the slit. In the intensity plot, variations in the light source have been removed by dividing by the intensity in the slit-jaw image and the result has been normalized.  The intensity is relatively constant along the slit, indicating that the nominal telescope position adequately centered the pupil in the spectrometer mirrors. The slit vignetting discussed in Section \ref{sec:alignspec} is apparent near the bottom of the intensity plot in Figure \ref{fig:tel-spec-align}d.

\begin{figure*}[htp]
	\centering
	\includegraphics[width=0.9\textwidth]{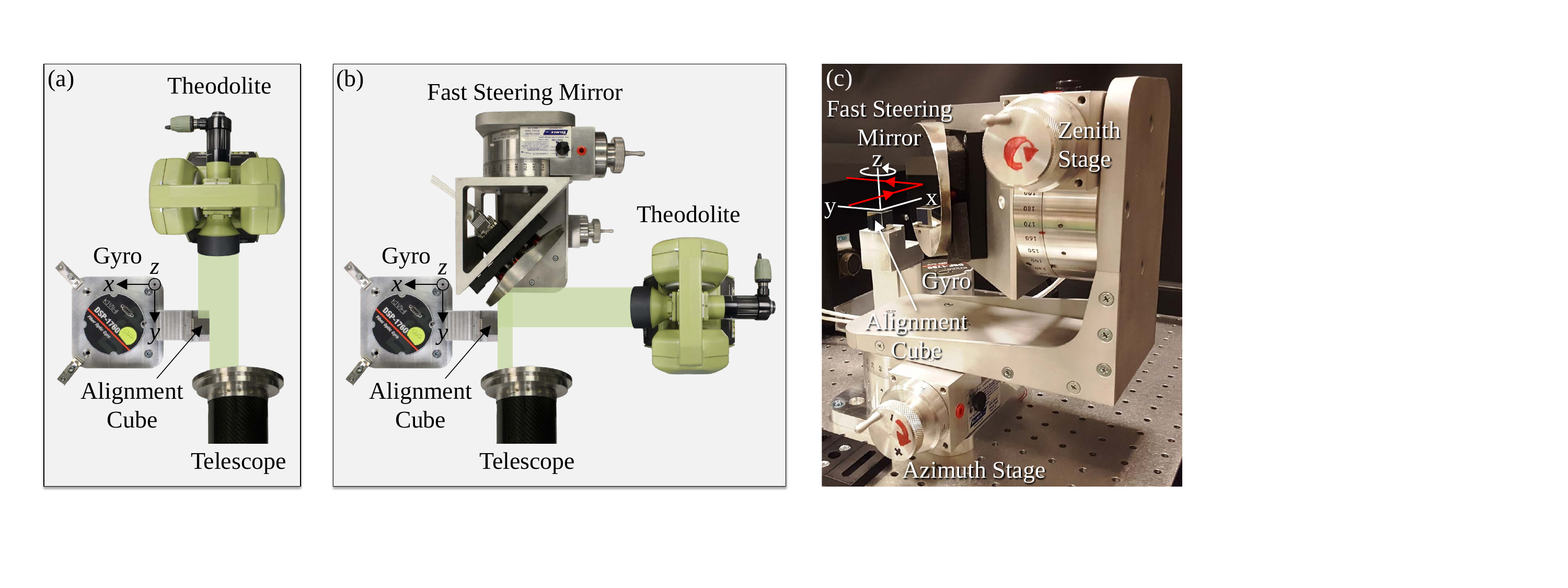}
	\caption{Alignment of the image stabilization system to the telescope. (a) Fiber-optic gyroscope to telescope. (b) Fast steering mirror to gyroscope and telescope. (c) Fast steering mirror adjustments.}
	\label{fig:FOG-FSM-align}
\end{figure*}

\subsubsection{Image Stabilization System Alignment}
In the final step of the optical alignment, the fiber-optic gyroscope and fast steering mirror were aligned to the telescope. The three gyroscope axes were referenced by a corner cube aligned to x arcmin using a coordinate measuring machine.  Two alignment holes in the bottom surface of the gyroscope package provided a precise reference for the gyroscope axes.

Figure \ref{fig:FOG-FSM-align}a shows the alignment of the gyroscope to the telescope.  The theodolite was aligned to the telescope line of sight by centering its crosshair on the center of the slit.  Part of the theodolite beam was retro-reflected by the gyro corner cube.  The gyroscope was tilted about its $x$ and $z$ axes until the $y$ axis was aligned to the theodolite, and therefore to the telescope line of sight. The final alignment error between the gyroscope and telescope was less than 10 arcsec in each axis. 

The fast steering mirror was installed and the theodolite was rotated 90 degrees to feed this mirror (Figure \ref{fig:FOG-FSM-align}b). The theodolite was aligned along gyro $x$ using the alignment cube.  Using the adjustments in Figure \ref{fig:FOG-FSM-align}c, the mirror assembly was rotated in azimuth and zenith until the image of the theodolite crosshair was centered at slit center. In this orientation, the fast steering mirror reflected the theodolite beam from gyro $x$ to gyro $y$ (the telescope line of sight), performing a 90 degree rotation about gyro $z$. The resulting azimuth and zenith stage positions defined the origin from which all future positions were measured.

\subsection{Eclipse Flight}
AIR-Spec observed the total eclipse from \mbox{GV~HIAPER} between 18:22 and 18:26 UTC. The observation took place over western Kentucky, near maximum duration.  The eclipse flight started and ended in Chattanooga, TN in order to minimize the fuel requirements and allow the Gulfstream V to fly at 14.3 km altitude, as high as possible above the IR-absorbing water vapor. To bring the spectrometer optics to thermal equilibrium, the instrument was cooled with liquid nitrogen beginning six hours before takeoff and ending 30 minutes before takeoff.

The flight path was designed to maximize the time spent in totality while optimizing the orientation of the aircraft before and during totality. Before entering the eclipse path, the GV flew a series of headings that compensated for the changing azimuth of the Sun.  This allowed the Sun to be acquired well before second contact with the instrument optimally positioned for totality.  The predetermined instrument position accounted for the likely wind direction, allowing the GV to fly along the eclipse path while remaining in an orientation that kept the total eclipse in the telescope.

\begin{figure}[htp]
	\centering
	\includegraphics[width=0.95\linewidth]{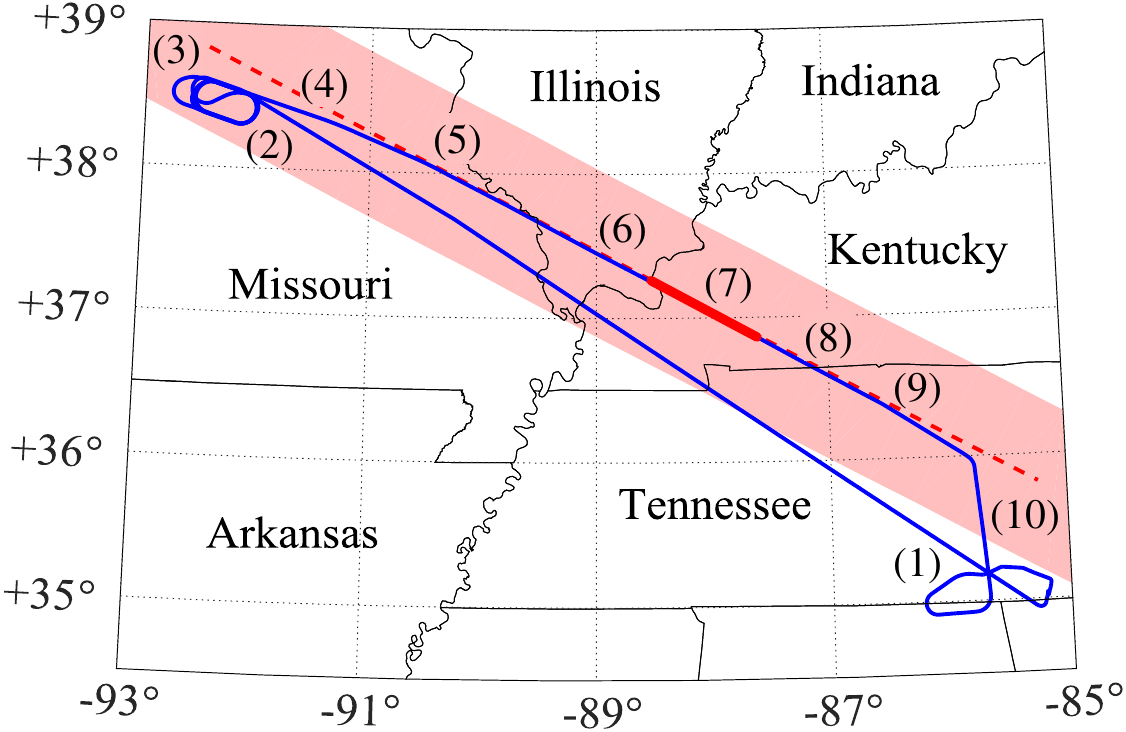}
	\caption{Eclipse flight and operations. The GV flight path is shown in blue, the totality centerline is marked by red dashes, and the totality swath appears in light red. The start time of each operation is noted with respect to second contact (C2) at 18:22 UTC.
	(1) Take off from Chattanooga, C2--130 min. 
	(2) Enter holding pattern, C2--80 min. 
	(3) Begin dry ice, C2--30 min. 
	(4) Acquire the Sun, C2--20 min. 
	(5) Set wavelength range, C2--10 min. 
	(6) Collect darks, C2--2 min.
	(7) Observe totality, C2--0 min.
	(8) Collect darks, C2+4 min. 
	(9) Collect calibration data, C2+10 min.
	(10) Land in Chattanooga, C2+45 min.}
	\label{fig:flight-path}
\end{figure}

Figure \ref{fig:flight-path} shows the three hour eclipse flight. The aircraft took off two hours before second contact (18:22 UTC), flew northwest for about 50 minutes, and entered a holding pattern.  The partial eclipse began, and the crew began adding dry ice to the IR camera interface.  About 25 minutes before totality, the GV exited the holding pattern and the Sun was acquired. The final spectral and spatial alignment was performed as described in Section \ref{sec:alignspec}.  AIR-Spec observed totality for four minutes, collecting dark frames before and after and photospheric calibration data near the end of the flight. The GV landed in Chattanooga about 45 minutes after totality.

\section{The August 21, 2017 Eclipse Data}
\label{sec:data}
The carefully planned flight and operations resulted in four minutes of data at four positions in the corona as well as a five second observation of the chromosphere \revis{(Table \ref{tab:obs-summary})}. All five target lines were observed (Table~\ref{tab:cor-lines}).  Prior measurement campaigns have confirmed the existence of \ion{Si}{10} at 1.431 \mic\ \citep{Penn1994, Kuhn1996}, \ion{S}{11} at 1.921 \mic\ \citep{Olsen1971,Kastner1993}, \ion{Mg}{8} at 3.028 \mic\ \citep{Munch1967, Olsen1971}, and \ion{Si}{9} at 3.935 \mic\ \citep{Kuhn1999, Judge2002}. AIR-Spec made the first detection of \ion{Fe}{9} at 2.853 \mic\ \citep{Samra2018}. Through coordinated observations with the Extreme Ultraviolet Imaging Spectrometer (EIS, \citealp{Culhane2007}), the AIR-Spec measurements provided insight into line excitation processes and coronal temperature and density \citep{Madsen2019}.

The post-processing scheme includes dark subtraction, defective pixel replacement, and image rectification and registration. Observations of the photosphere are used to calibrate wavelength, pointing, plate scale, throughput, and resolution.  

\subsection{Summary of Observations}
AIR-Spec observed the corona at four positions, sampling the west limb, a solar prominence, and the east limb (including an active region). A fifth position measured the chromosphere at third contact. Table~\ref{tab:obs-summary} lists the time spent at each position and the number of frames acquired in each camera.  Most of totality was spent on the west limb to coordinate with other observatories.

\begin{deluxetable}{lccc}[!htp]
    \tablecaption{Summary of the total eclipse observations.}
    \label{tab:obs-summary}
    \tablehead{\colhead{Slit Position} & \colhead{Duration} & \colhead{IR Frames} & \colhead{SJ Frames}}
    \startdata
    1. West limb & 63.5 sec & \phn953 & 2083\\
    2. Prominence & 41.5 sec & \phn622 & 1359\\
    3. East limb & 35.7 sec & \phn536 & 1171\\
    4. Prominence & \multirow{2}{*}{81.3 sec} & \multirow{2}{*}{1219} & \multirow{2}{*}{2668}\\
    \phantom{4. }\& west limb & & & \\
    5. Chromosphere & \phn5.0 sec & \phn\phn75 & \phn164\\
    \enddata
\end{deluxetable}

\begin{figure*}[htp]
	\centering
	\includegraphics[angle=0,width=1\textwidth]{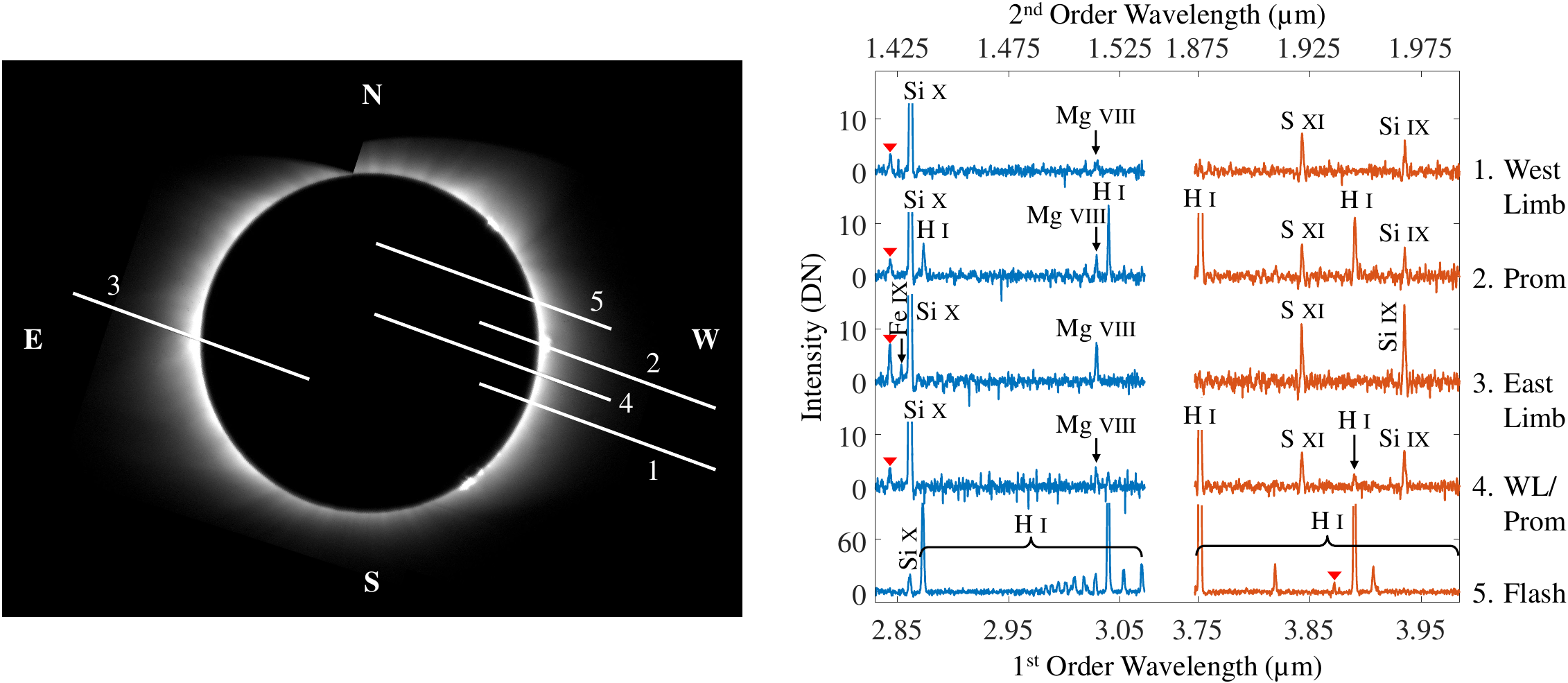}
	\caption{Overview of total eclipse observations. Slit positions superimposed on slit-jaw image (left) and average IR spectra at each slit position (right). \revis{Red triangles mark the spectral artifacts.}}
	\label{fig:obs-overview}
\end{figure*}

Figure \ref{fig:obs-overview} shows the average slit position for each observation superimposed on a composite of images from the slit-jaw camera (left) and the mean infrared spectrum corresponding to each slit position (right).  Each baseline-subtracted spectrum was averaged completely in time and over 35 arcsec at the limb.  Neutral hydrogen was observed in the prominence and chromosphere. Features appearing at 2.843 \mic\ in positions 1--4 and  3.021, 2.854, and 3.873 \mic\ in position 5 are not emission lines but artifacts caused by stray reflections inside the slit-jaw substrate \citep{Samra2019}.

\begin{figure*}[htp]
	\centering
	\includegraphics[angle=0, width=0.98\textwidth]{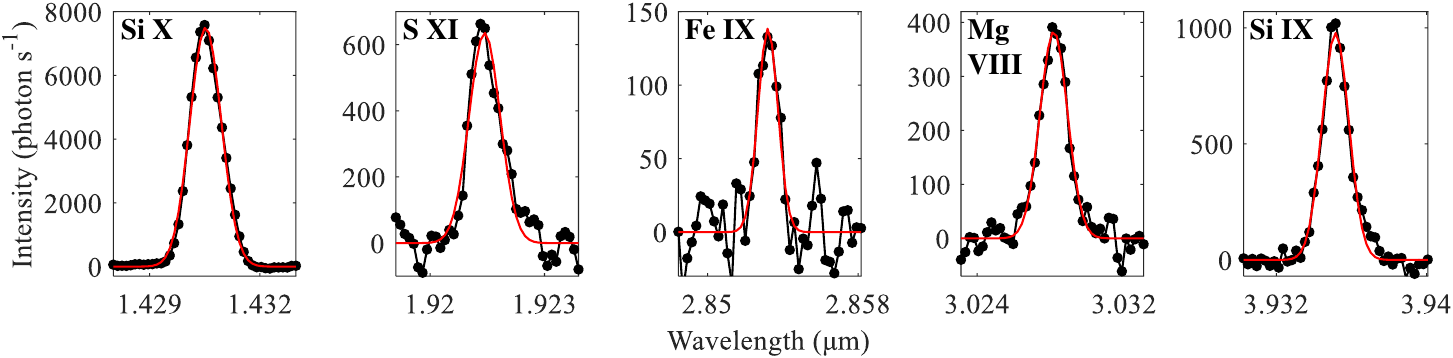}
	\caption{Gaussian fits to the coronal lines in position 3 (east limb active region). The measurements were averaged in time and over the 21 pixels (49 arcsec) nearest the lunar limb, corresponding to an average radius of 1.05 \Rs.}
	\label{fig:fits}
\end{figure*}

\ion{Fe}{9} was observed in position 3 (east limb), and the other four target lines were observed in all of the coronal positions. Gaussian fits to each coronal line (Figure \ref{fig:fits}) provided the center wavelength in positions 1--4 (Table~\ref{tab:cor-lines}). Uncertainties in the wavelengths of the observed coronal lines arise from a combination of the error in the Gaussian fits and uncertainties in the wavelength mapping (Table~\ref{tab:wave-cal}).  

\begin{deluxetable*}{rccccc}
\tablecaption{Measured wavelengths for the AIR-Spec coronal lines.}
\label{tab:cor-lines}
\tablehead{& \colhead{\ion{Si}{10}} & \colhead{\ion{S}{11}} & \colhead{\ion{Fe}{9}} & \colhead{\ion{Mg}{8}} & \colhead{\ion{Si}{9}}}
\startdata 
1. & $14304.4 \pm 2.1$ \AA & $19214.4 \pm 0.7$ \AA &                       & $30278.4 \pm 3.2$ \AA & $39351.8 \pm 1.2$ \AA \\ 
2. & $14304.1 \pm 2.0$ \AA & $19213.2 \pm 0.7$ \AA &                       & $30280.2 \pm 2.8$ \AA & $39353.0 \pm 1.2$ \AA \\ 
3. & $14305.4 \pm 1.5$ \AA & $19214.4 \pm 0.5$ \AA & $28532.3 \pm 3.1$ \AA & $30281.9 \pm 2.0$ \AA & $39351.8 \pm 0.8$ \AA \\ 
4. & $14305.0 \pm 1.8$ \AA & $19213.8 \pm 0.6$ \AA &                       & $30280.5 \pm 2.7$ \AA & $39353.7 \pm 1.0$ \AA \\[2pt]
\enddata
\tablecomments{Wavelength estimates and standard errors are reported for each of the four coronal slit positions.  The \ion{Fe}{9} line was only seen in position 3 \citep{Samra2018,Samra2019}. All wavelengths are given in vacuo and have been corrected for solar rotation.}
\end{deluxetable*}

\subsection{Image Processing}
Each infrared camera frame was processed to remove the dark background, replace defective pixels, align the spatial and spectral axes with the image, register to other frames along the slit, and remove interference fringes from the 4 \mic\ channel. \revis{Fringes were not observed in the 3 \mic\ channel.} Two-dimensional image registration was performed on the slit-jaw images.

\subsubsection{Dark Subtraction}
Because of the significant thermal background discussed in Section \ref{sec:bgreduction}, it was necessary to subtract a dark frame from each infrared data frame in order to see the IR corona. \revis{Due to the high spatial uniformity of the focal plane (Figure \ref{fig:nuc}), the weakness of the coronal signal compared to the background, and the difficulty of taking flat field data with the spectrometer at temperature, dark subtraction was used in lieu of a flat field correction. The dark subtraction removes the offset non-uniformity (Figure \ref{fig:nuc}, first panel) and any gain non-uniformity (second panel) associated with the background signal. It cannot remove the gain non-uniformity from the coronal signal, but this residual is insignificant compared to photon noise across the line. The local gain variation of $<0.02$ DN/DN (third panel) corresponds to only a few DN on the combined signal from the line and continuum.}

\begin{figure}[htp]
	\centering
	\includegraphics[width=0.9\linewidth]{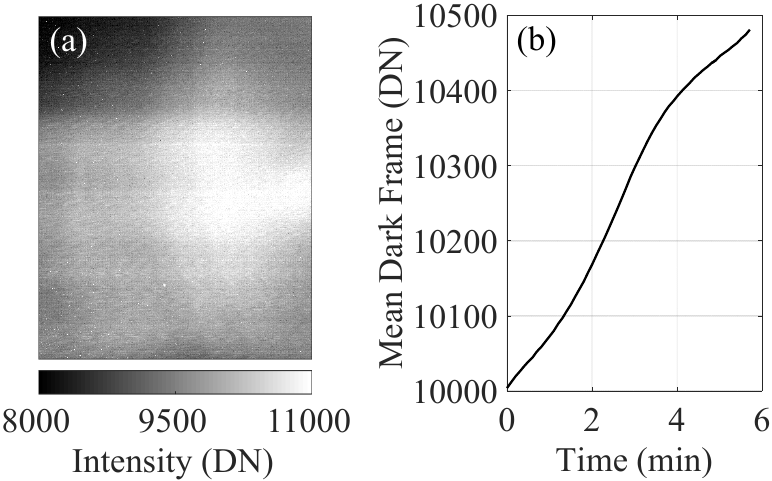}
	\caption{Behavior of the dark background \revis{at 60 ms exposure time.} (a) Dark frame at the end of totality. (b) Six minute time series of the spatial average.}
	\label{fig:dark-frame-time}
\end{figure}

Dark frames were collected before and after totality by blocking light from the telescope at the vacuum chamber entrance window. Figure~\ref{fig:dark-frame-time} shows the dark frame immediately after totality (a) and the time variation of the mean dark frame over the 6 minutes following totality (b). The background image has a mean of about 10,000~DN and significant structure resulting from bright regions inside the spectrometer.  In comparison, the measured intensity of the brightest coronal line is about 150 DN.  The background increases with time at a rate of about 80 DN/minute as the liquid nitrogen evaporates and the dry ice sublimates. 

\begin{figure*}[htp]
	\centering
	\includegraphics[width=0.85\textwidth]{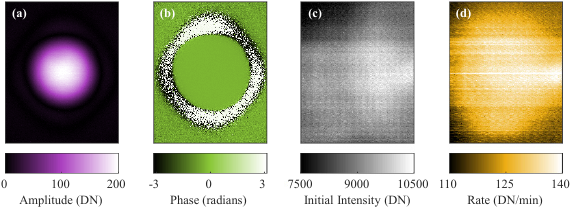}
	\caption{Fit parameters for the time evolution of the dark background. (a) Initial intensity in DN. (b) Rate of change in DN/minute. (c) Sine wave amplitude in DN. (d) Sine wave phase in radians.}
	\label{fig:dark-parameters}
\end{figure*}

For pixels near the center of the array, a 4.2 minute sinusoidal trend is superimposed on the warm-up.  The sine wave comes from the Stirling cooler used to chill the detector.  The cold piston of the cooler is in contact with the middle of the focal plane and moves toward and away from the focal plane at 60 Hz, the operating frequency of the cooler. This generates a time-varying acceleration in the pixels near the center of the detector, introducing a piezoelectric effect that changes the intensity in those pixels.  The 4.2 minute period is a result of aliasing the 60 Hz fluctuation by sampling at the 15 Hz frame rate. The amplitude $A$ and phase $\phi$ of the aliased sine wave are given by 
\begin{align}
	A&=A_0\sin\left[\pi\;\textrm{frac}\left(T_{exp}\cdot60\textrm{ Hz}\right)\right]\label{eq:amp}\\
	\phi&=\phi_0+\pi\;\textrm{frac}\left(T_{exp}\cdot60\textrm{ Hz}\right)\label{eq:phase}
\end{align}
where ``frac'' denotes the fractional part of the argument (i.e. the argument modulo 1), $T_{exp}$ is the exposure time, and $A_0$ and $\phi_0$ are the amplitude and phase of the original (unaliased) 60 Hz wave. $A_0$ and $\phi_0$ vary from pixel to pixel but remain constant in time.  They were measured after the eclipse and are shown in panels (a) and (b) of Figure \ref{fig:dark-parameters}.

Because the background varied significantly over the four minutes of totality, its time evolution was modeled using an eight minute series of dark frames from before and after totality. 90 frames (6 seconds) of data were averaged to produce each sample in the time series. The signal in each pixel was assumed to have the form
\begin{equation}
\begin{split}
I(t)=I_0 + r(t-t_0) &+ a(t-t_0)^2\\
&+A\sin\left(\frac{2\pi (t-t_0)}{\rm{4.2\ min}} + \phi\right), \label{eq:dark-evol}
\end{split}
\end{equation}
where $I$ is the pixel intensity in DN, $t$ is the time each frame was collected, $t_0$ is an arbitrary start time, $I_0$ is the initial intensity in DN, $r$ is the linear rate of change in DN/minute, $a$ is the acceleration in DN/minute$^2$, $A$ is the sine wave amplitude in DN, and $\phi$ is the sine wave phase in radians. $A$ and $\phi$ were calculated using Equations \ref{eq:amp} and \ref{eq:phase}, and $t_0$ was selected such that the observed phase of the dark time series was equal to $\phi$ at $t=t_0$. The sinusoidal term was subtracted from the time series and $I_0$, $r$, and $a$ were estimated for each pixel using a least-squares fit.  Maps of $I_0$ and $r$ are shown in panels (c) and (d) of Figure \ref{fig:dark-parameters}.

\subsubsection{Defective Pixel Replacement}
\begin{figure*}[htp]
	\centering
	\includegraphics[angle=0,width=0.85\textwidth]{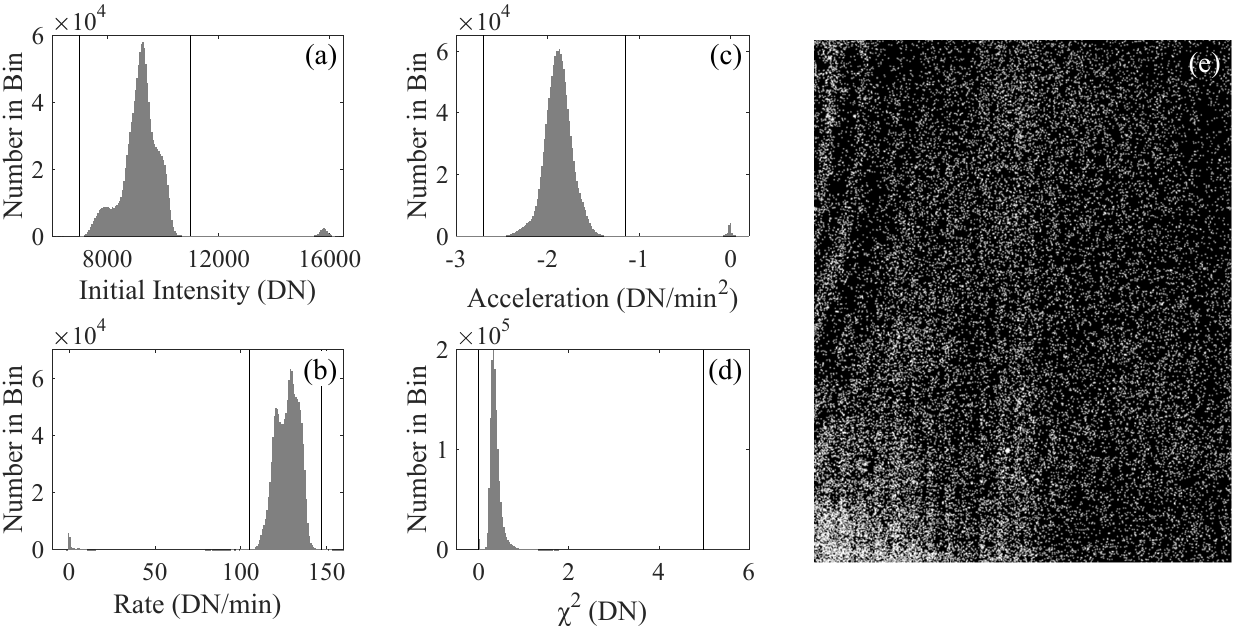}
	\caption{Defective pixels. Histograms are used to identify outliers in initial intensity (a), rate (b), acceleration (c), and goodness of fit (d).  Good pixels lie between the black vertical lines. Defective pixel map (e). Bad pixels are shown in white and make up 2\% of the array.}
	\label{fig:bad-pix}
\end{figure*}

At 60 ms exposure time, defective pixels made up about 2\% of the focal plane array. Bad pixels were identified as outliers in the distributions of initial intensity, rate, or acceleration or as pixels with poor fits to Equation \ref{eq:dark-evol}. They were replaced using bilinear interpolation. Figure \ref{fig:bad-pix} shows the histograms for initial intensity (a), rate (b), acceleration (c), and $\chi^2$ (d), with black vertical lines marking the cutoff values for good pixels.  Defective pixels are shown in white in Figure \ref{fig:bad-pix}e. 

\subsubsection{Geometric Correction}
\label{sec:geocorr}
\begin{figure*}[htp]
	\centering
	\includegraphics[width=1\textwidth]{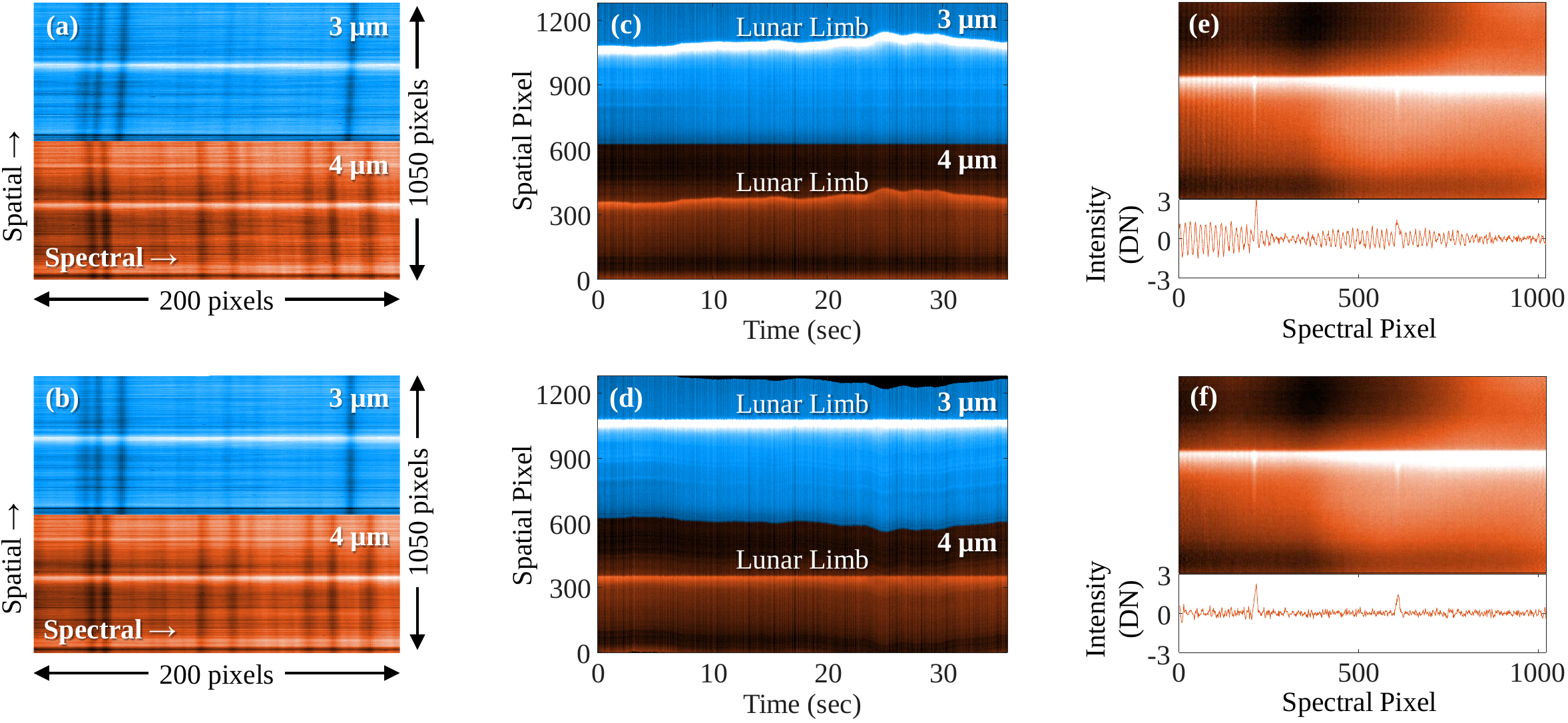}
	\caption{IR image processing: geometric correction, image registration, and fringe removal. The \revis{3.5$^\circ$ angle between the slit and the grating grooves} shears spectral and spatial features with respect to the detector (a).  After geometric correction, the spectral and spatial axes are aligned with the detector axes in both channels (b).  Due to pointing jitter, the lunar limb crosses the slit at a different place in each frame (c). The high-contrast limb is used to detect the jitter, and each frame is shifted an integer number of pixels to align with the others (d). In the 4 \mic\ channel, interference fringes obscure the emission lines (e). The fringes are removed using a notch filter at the fringe frequency (f).}
	\label{fig:geom-align-fringe}
\end{figure*}

After dark subtraction and defective pixel replacement, a geometric correction was applied to remove spectral and spatial shear in the infrared images.  \revis{Shear arose from the fact that the slit was not parallel to the grating grooves, by design. (The grating was tilted 3.5$^\circ$ to the slit in the cross-dispersion direction in order to send light to the focus mirrors.)} The shear is apparent in Figure \ref{fig:geom-align-fringe}a, which shows IR data from scattered photospheric light just before totality.  The structure along the spectral dimension comes from photospheric and atmospheric absorption, while the spatial structure comes from dust on the slit and variations in the scattered intensity. The spectral shear is more apparent in the image because that dimension is magnified.

Shear was removed by computing a coordinate transformation to align the image axes with the sheared features and then interpolating the data from sheared coordinates to Cartesian coordinates. A gradient filter was used to detect lines along the spectral and spatial features. Each line was defined by two $(x,y)$ coordinate pairs, which were mapped to corrected coordinates $(u,v)$ corresponding to horizontal and vertical lines.  Because the sheared lines were straight and parallel, the $2\times3$ affine transformation $T$ was used to map each $(x,y)$ to $(u,v)$ according to
\begin{equation}
\label{eq:affine}
\begin{bmatrix}
u \\
v
\end{bmatrix}
=
T
\begin{bmatrix}
x \\
y \\
1
\end{bmatrix}
\end{equation}
A least squares estimate for $T$ is given by
\begin{equation}
\hat T=UX^T\left(XX^T\right)^{-1},
\end{equation}
where  $U=
\begin{bmatrix}
u_1 & u_2 & \dots & u_n \\
v_1 & v_2 & \dots & v_n
\end{bmatrix} \rm{,\ }
X=\begin{bmatrix}
x_1 & x_2 & \dots & x_n \\
y_1 & y_2 & \dots & y_n \\
1 & 1 & \dots & 1
\end{bmatrix}$, and $n\geq3$ is the number of features (lines) extracted.

$\hat T$ was used to produce a sheared grid aligned with the spectral and spatial features, and the IR data was mapped from the sheared grid to a Cartesian grid using bilinear interpolation. Separate transformations were used to correct the 3 \mic\ and 4 \mic\ channels. Figure~\ref{fig:geom-align-fringe}b shows the IR data after geometric correction.  The spectral and spatial features are aligned with the image axes. 

\subsubsection{Image Registration} \label{sec:coalign}
\begin{figure}[htp]
	\centering
	\includegraphics[angle=0,width = 0.95\linewidth]{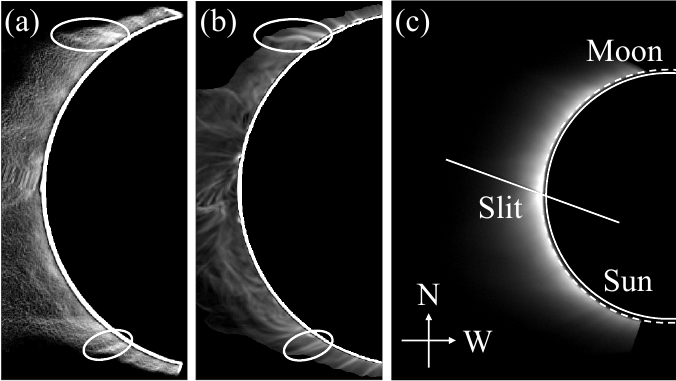}
	\caption{Slit-jaw image registration.  By matching features in AIR-Spec (a) and AIA 193 (b), the slit-jaw frames are rotated to orient solar north up. Moon center is found by fitting the lunar limb to a circle (c). Each frame is shifted horizontally and vertically to place Moon center in the same location.}
	\label{fig:sj-align} 
\end{figure}

In order to remove residual pointing jitter, image registration was performed on frames from both cameras.  The slit-jaw images were registered in $x$, $y$, and rotation to place solar north up and Moon center at the same location in all frames (Figure \ref{fig:sj-align}).  The rotation angle was found by comparing the IR corona to the 193 \AA\ channel of the Atmospheric Imaging Assembly (AIA, \citealp{Lemen2012}), after enhancing features by applying a radial filter and computing the image gradients. Moon center was found by fitting the limb of the Moon to a circle.

The IR images were registered along the spatial dimension by placing the limb of the Moon at the same pixel in all frames. First, each frame was summed along the spectral dimension to produce the integrated intensity at every pixel along the slit. Figure \ref{fig:geom-align-fringe}c shows the spectrally integrated intensity as a function of spatial pixel and time. The inner corona appears as a wavy bright line in each channel, with a sharp intensity change at the limb of the Moon.  The spectrally integrated intensity was differentiated with respect to the distance from Sun center, and the pixel with the largest derivative was defined as the limb of the Moon. Each frame was shifted an integer number of spatial pixels to align the location of the limb. Figure \ref{fig:geom-align-fringe}d shows the spectrally integrated intensity after alignment. 

\subsubsection{Fringe Removal}
\label{sec:fringe}
Interference fringes appeared in the 4 \mic\ channel of the infrared data, due to an etalon effect in the four uncoated sapphire surfaces of the aircraft window. The fringes were oriented vertically on the image and had a period of about 14 pixels (170 \mic) and a peak-to-peak amplitude of 5--10 DN, \revis{ equivalent to about \mbox{1.5--3}~\textmu \Bs\ at a wavelength of 3.935 \mic.} Figure \ref{fig:geom-align-fringe}e shows a sample 4~\mic\ image and its spatial (row) mean.  The fringes are bright enough to obscure the two emission lines in the image.

\begin{figure}[htp]
	\centering
	\includegraphics[width=1\linewidth]{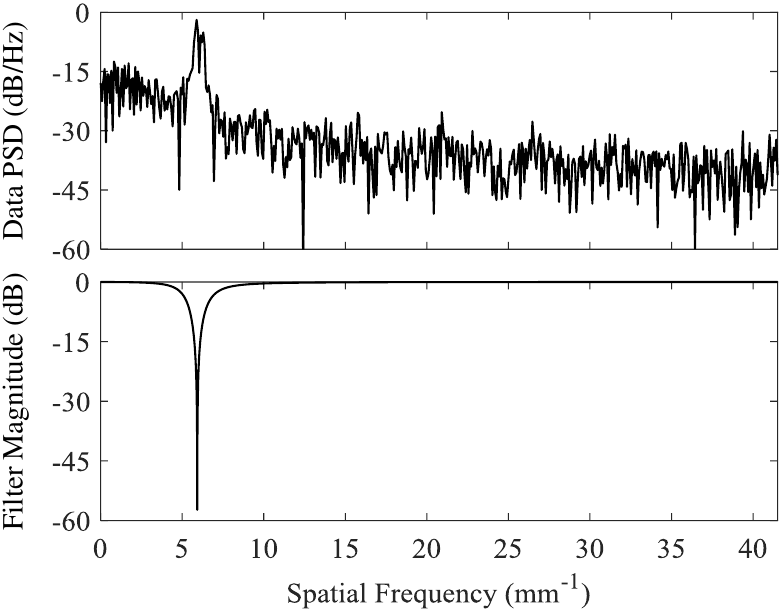}
	\caption{Power spectral density of the mean row in the 4 \mic\ channel (top). Fringes with a period of 14 pixels (170 \mic) appear as a peak at 6 mm$^{-1}$. The notch filter (bottom) attenuates frequencies between 5 and 7 mm$^{-1}$, removing the fringes.}
	\label{fig:notch-filter}
\end{figure}

The fringes can be removed by applying a notch filter at the fringe frequency. Figure \ref{fig:notch-filter} shows the power spectral density (blue) of the mean row in the 4 \mic\ channel. A peak appears at the fringe frequency of 6 mm$^{-1}$. The notch filter (black) attenuates frequencies between 5 and 7 mm$^{-1}$.  Figure \ref{fig:geom-align-fringe}f shows the sample 4 \mic\ image after fringe removal.  The emission lines are clearly visible. The fringes were not removed in the published data.

\subsection{Calibration}
The AIR-Spec wavelength, plate scale, pointing, and throughput were calibrated to map spectral pixel to wavelength, spatial pixel to solar coordinates, and image intensity to spectral radiance.  A calibration of the PSF provided a measurement of the spectral and spatial instrument resolution.

\subsubsection{Wavelength Calibration} \label{sec:wavecal}

AIR-Spec has a linear mapping from spectral pixel to wavelength. The mapping is constant with respect to spatial pixel (after geometric correction), but it varies in time by more than 4 spectral pixels over the course of the eclipse observation. The start wavelength drifts due to thermal changes and jitters due to vibration from the aircraft and camera chiller, while the dispersion remains fixed in time.  The temporal trends are different in the two channels, suggesting that the two focus mirrors are the main source of the variations. 

These temporal variations in start wavelength can be calibrated as long as they happen on timescales longer than one exposure. This is the case for the thermal drift and the aircraft-related jitter. High-frequency vibration, likely due to the camera chiller, cannot be removed and causes a double peak or broad peak in many of the line observations, especially in the 3 \mic\ channel. This \revis{effect appears because the tip/tilt mounts for the focus mirrors are not sufficiently constrained. It}  reduces the effective spectral resolution and increases the uncertainty in the line wavelengths in Table \ref{tab:cor-lines}.

Wavelength calibration took place in two steps: first the absolute time-averaged wavelength was calibrated for each slit position, and then the wavelength axis was shifted at each time step to account for the variation in start wavelength. In the coronal slit positions (1--4), the average wavelength mapping was estimated by comparing AIR-Spec's measurement of telluric absorption in the scattered continuum with the ATRAN atmospheric model \citep{Lord1992}, available online at \url{https://atran.arc.nasa.gov}. The modeled atmosphere was smoothed to the AIR-Spec resolution and a nonlinear curve fitting routine was used to stretch and shift the measured data to achieve the best match to the model. In the chromospheric observation (position~5), the time-averaged wavelength was estimated by fitting the hydrogen lines with Gaussians and then doing a linear fit from center pixel to rest wavelength. Table \ref{tab:wave-cal} lists the time-averaged wavelength mapping coefficients and their uncertainties for slit position 3 (east limb). 

\begin{deluxetable}{ccc}
\tablecaption{Wavelength calibration coefficients and standard errors in slit position 3.}
\label{tab:wave-cal}
\tablehead{\colhead{Channel} & \colhead{Start Wavelength (\AA)} & \colhead{Dispersion (\AA/pixel)}} 
\startdata
1.5 \mic & $15359.38\pm0.98$& $-1.18338\pm0.00123$\\
\phantom{1.}2 \mic & $19926.94\pm0.35$ & $-1.16811\pm0.00048$\\
\phantom{1.}3 \mic & $30718.75\pm1.97$ & $-2.36675\pm0.00246$\\
\phantom{1.}4 \mic & $39853.88\pm0.71$ & $-2.33622\pm0.00097$\\
\enddata
\end{deluxetable}

\begin{figure}[htp]
	\centering
	\includegraphics[angle=0,width=1\linewidth]{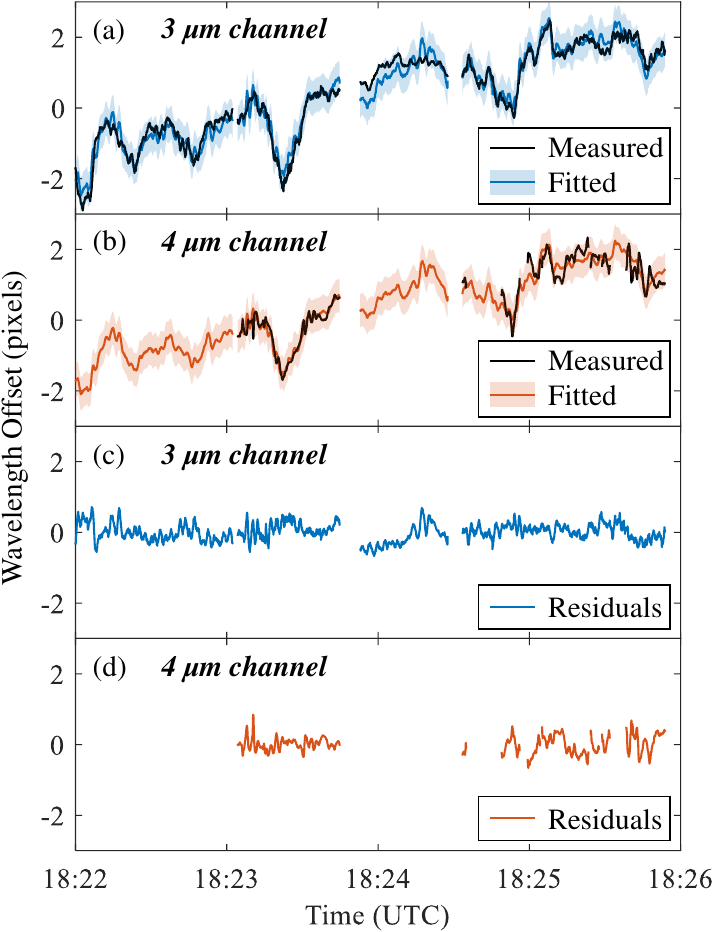}
	\caption{Modeling wavelength drift/jitter as a linear combination of attitude and time. Measurements and fit, 3~\mic\ channel (a) and 4~\mic\ channel (b). The shaded regions are the 95\% prediction intervals. Fit residuals, 3~\mic\ channel (c) and 4~\mic\ channel (d).}
	\label{fig:wave-cal-time} 
\end{figure}

After estimating the average wavelength axis for each slit position, the strong lines \ion{Si}{10} 1.431 \mic\ and \ion{H}{1} 1.876~\mic\ were used as tracers for the time-varying wavelength offset in the 3 \mic\ and 4 \mic\ channels. The line measurements were summed to 0.9 s effective exposure time and fit with Gaussians to provide the line centers as a function of time. Each line center time series was modeled as a linear combination of yaw, pitch, roll, their temporal derivatives, and time, resulting in a residual offset of less than 1 pixel over the duration of the total eclipse (Figure \ref{fig:wave-cal-time}). These models were then used to predict the 3 \mic\ and 4 \mic\ start wavelength for every frame.

\subsubsection{Plate Scale Measurement and Pointing Calibration} \label{sec:platescale}
\begin{figure}[htp]
	\centering
	\includegraphics[width = 1\linewidth]{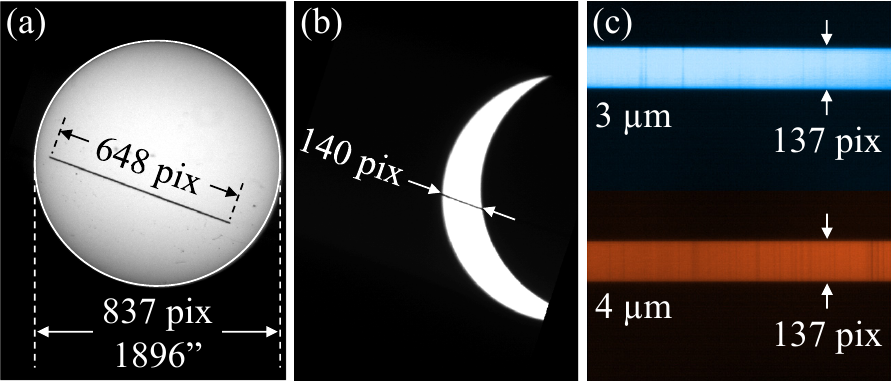}
	\caption{Measuring plate scale.  The slit-jaw camera plate scale and slit length are calculated experimentally by fitting the limb of the solar disk to measure its diameter (a). The plate scale is 2.26 arcsec/pixel, and the slit length is 1.55~\Rs. The IR camera plate scale is computed by comparing simultaneous observations of the partial eclipse in the slit-jaw camera (b) and IR camera (c). The IR camera has a plate scale of 2.31 arcsec/pixel, 2\% larger than the slit-jaw camera.}
	\label{fig:plate-scale} 
\end{figure}

The slit-jaw camera plate scale and slit length were measured using a solar observation from two days before the eclipse. The solar diameter was measured by fitting the limb of the disk (Figure \ref{fig:plate-scale}a), and the plate scale and slit length were found to be 2.26 arcsec/pixel and 1.55\Rs. The effective slit length was 10\% shorter because the slit-jaw mount clipped light from the slit, as mentioned in Section \ref{sec:alignspec}. 

The IR camera plate scale was computed by comparing simultaneous observations of the partial eclipse in the slit-jaw camera (Figure \ref{fig:plate-scale}b) and IR camera (Figure \ref{fig:plate-scale}c). From the width of the  intensity profiles along the slit, the IR plate scale was found to be 2\% larger than that of the slit-jaw camera. The IR camera achieved the designed plate scale of 2.31 arcsec/pixel, while the slit-jaw camera was restricted by the range on its translation stage, as described in Section \ref{sec:alignsjcamera}.

The plate scales were used to map spatial pixel to Helioprojective-Cartesian coordinates in both sets of images.  Once the slit-jaw frames were aligned with respect to Moon center as described in Section \ref{sec:coalign}, the solar coordinates of each pixel were computed by accounting for the drift of the Moon relative to the Sun. The time-varying offset between Sun center and Moon center was computed by cross-correlating on a solar prominence over the four minute observation. In each IR image, the spatial pixel of the lunar limb crossing was determined using the method explained in Section \ref{sec:coalign}.  The limb crossing was mapped to solar coordinates using the corresponding slit-jaw image, and the coordinates for the other spatial pixels were determined using the IR plate scale. 

\subsubsection{Throughput Calibration}

\begin{figure}[htp]
	\centering
	\includegraphics[width=1\linewidth]{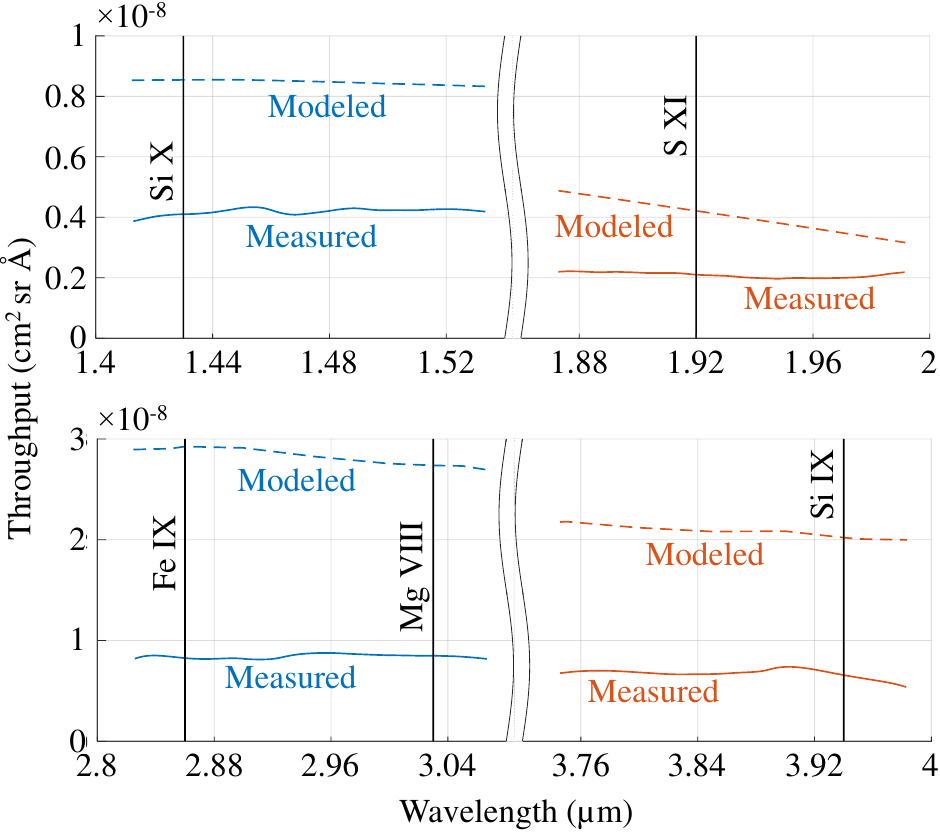}
	\caption{\revis{Measured} and modeled throughput.}
	\label{fig:throughput} 
\end{figure}

We define throughput as the product of overall optical efficiency, geometric area \revis{of the telescope entrance aperture}, slit width in wavelength units, and solid angle subtended by a pixel. The AIR-Spec throughput was estimated using both a bottom-up model and a top-down measurement.  The forward model \revis{(dashed lines in Figure \ref{fig:throughput})} was computed by simply multiplying the slit width, solid angle, and geometric area from the optical design by the estimated efficiency of each element in the optical train, from the aircraft window to the bandpass filter immediately in front of the detector.  \revis{We estimate a total optical efficiency (without QE) of about 20\% in the 1.5 \mic\ channel, 10\% in the 2 \mic\ channel, 35\% in the 3 \mic\ channel, and 25\% in the 4 \mic\ channel. The grating (20--80\% reflectivity) and the aircraft viewport (70\% transmission) are the largest individual contributors to the efficiency estimate.}

The top-down throughput calibration relied on photospheric measurements acquired during a test flight four days before the eclipse. In order to separate the contributions from first and second order, bandpass filters were used to isolate light near 1.5, 2, 3, and 4 \mic. All observations were acquired with a broadband solar filter (\revis{transmission} $\approx10^{-3}$) in place. The AIR-Spec data were compared with the expected spectrum at the entrance to the instrument, which was taken as the product of the radiance from the photosphere, the transmission through earth's atmosphere, and the transmission through the bandpass and solar filters. The photospheric library spectrum came from a combination of ground-based \citep{Wallace1996} and space-based \citep{Farmer1989} measurements and a model from R. Kurucz (\url{http://kurucz.harvard.edu/stars/sun}). The ATRAN model \citep{Lord1992} was used to provide an estimate of the atmospheric transmission. The transmission of each filter was measured by J. Hannigan of NCAR Atmospheric Chemistry Observations and Modeling \revis{using a laboratory Fourier transform spectrometer (FTS)}. 

\begin{figure}[htp]
	\centering
	\includegraphics[angle=0,width=1\linewidth]{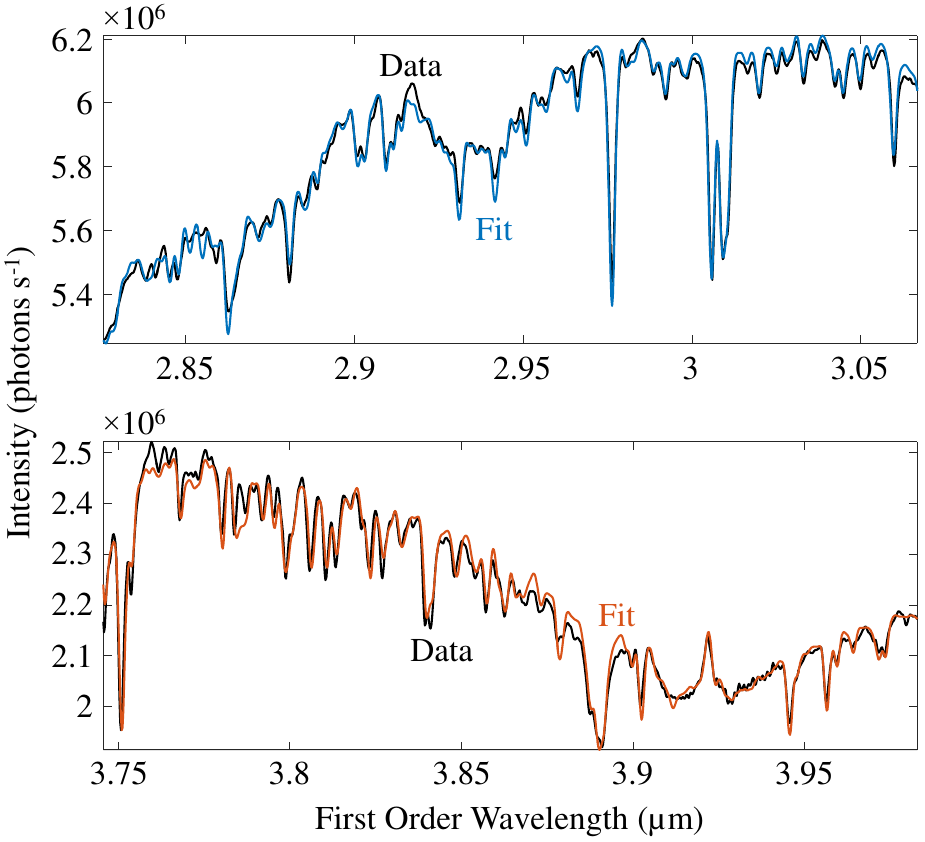}
	\caption{Checking the inverse-modeled throughput. The black line is AIR-Spec data, converted from DN \revis{(1 DN = 3.25 e$^-$)} to photon~s$^{-1}$. The red line shows the photospheric library spectrum after multiplication by the inverse-modeled throughput.}
	\label{fig:throughput-fit} 
\end{figure}

The throughput was measured as a function of wavelength by fitting a smoothing spline to the quotient of the measured spectrum and library spectrum. The result is shown in the solid lines in Figure \ref{fig:throughput}. Figure \ref{fig:throughput-fit} shows the agreement between the AIR-Spec data and library spectrum after the latter was multiplied by the throughput and first and second order were summed together.

The measured throughput estimate is 2--3 times smaller than the modeled estimate (Figure \ref{fig:throughput}). We are still exploring possible causes for this discrepancy, but we suspect a combination of factors. \revis{(1)}~The geometric area of the instrument was probably \revis{slightly} smaller than designed due to the angle of the Sun on the feed mirror and misalignments inside the spectrometer. \revis{(2)~When the highly attenuating solar filter was placed into the FTS path to measure its transmission, it was necessary to increase the aperture diameter and preamplifier gain to produce a measurable signal at the FTS detector. The change in settings adds some uncertainty to the transmission measurement. (3)} The \revis{photospheric} AIR-Spec measurements may have been influenced by limb darkening, in which case the library spectrum would have overestimated the intensity of the photosphere. \revis{(4) For optical elements without transmission/reflectivity measurements, we based our efficiency estimates on the typical performance specified by the manufacturer.  This likely  resulted in an overestimate of the overall efficiency.} In the future, we will use a calibrated lamp to measure throughput in the laboratory, where we have fewer sources of error and greater control over the experimental conditions.
	
\revis{Based on our understanding of the AIR-Spec alignment and the linearity of the FTS used to measure filter transmission, we believe the first two factors together are responsible for $<$20\% of the  measured-modeled throughput discrepancy. In addition, these errors cannot explain the wavelength dependence of the discrepancy. We also expect limb darkening to contribute no more than 20\% of the difference, as the average intensity across the solar disk is about 80\% of the intensity at the center at 500~nm wavelength.  We believe at least 50\% of the discrepancy can be explained by overestimates of the individual optical efficiencies, especially the reflectivity of the protected-silver coating on the fast-steering mirror. This coating degraded significantly between delivery and flight, but the mirror was an integrated unit that could not be easily recoated without impact to the tight schedule.} 

\subsubsection{PSF Measurement} \label{sec:psf}

\begin{figure}[htp]
	\centering
	\includegraphics[angle=0,width=0.85\linewidth]{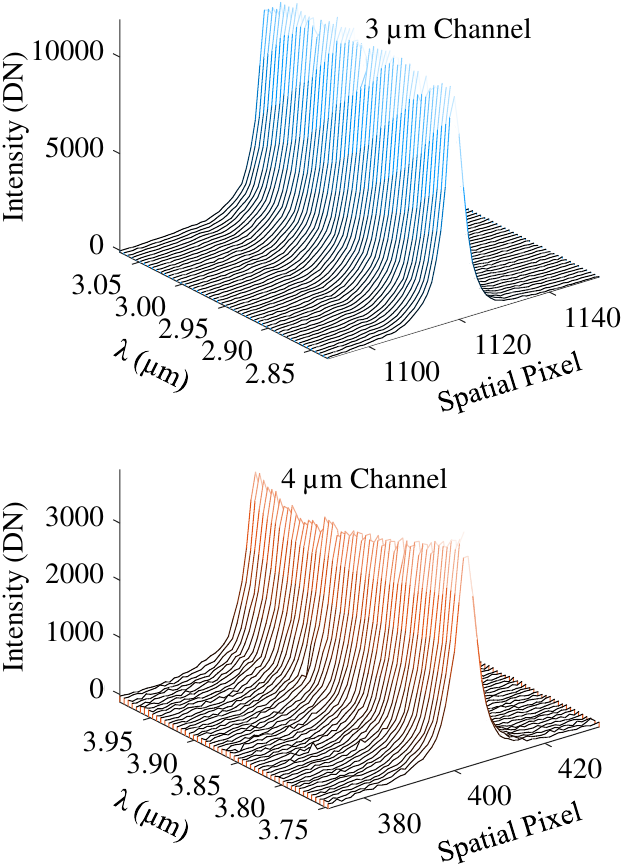}
	\caption{Measuring point-spread function.}
	\label{fig:psf} 
\end{figure}

The AIR-Spec point-spread function (PSF) was measured using an observation of the photosphere immediately after third contact.  At the time of the observation, the exposed photosphere was a bright point source, and single frame at an exposure time of 0.3 ms provided a high-SNR measurement of the resolution in both channels. The raw data are shown in Figure \ref{fig:psf}. The photosphere appears as a bright line spanning all wavelengths but only a few spatial pixels.

\begin{figure}[htp]
	\centering
	\includegraphics[angle=0,width=1\linewidth]{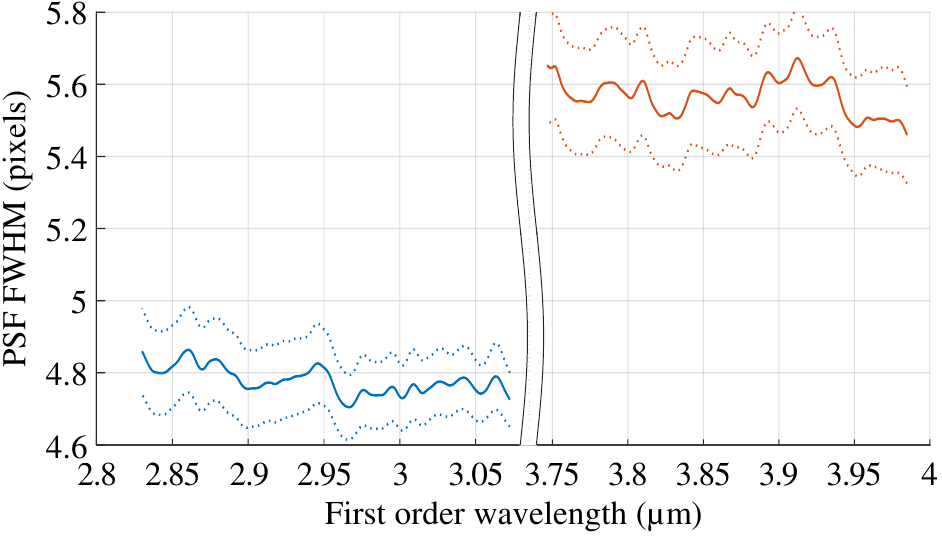}
	\caption{Point-spread function FWHM across the detector. The dotted lines represent 95\% confidence intervals.}
	\label{fig:psf-fwhm} 
\end{figure}

The PSF full width at half maximum (FWHM) was measured by fitting a Lorentzian to the data in each spectral channel. The FWHM is plotted as a function of wavelength in Figure \ref{fig:psf-fwhm}. Due to the optical design, the point-spread function is about 1 pixel broader in the 4 \mic\ channel than in the 3 \mic\ channel. The FWHM of 4.8--5.6 pixels corresponds to a spatial resolution of 11--13 arcsec and a spectral resolution of about 14--15~\AA\ in first order (half that value in second order), once the effect of the slit is included. Measurements of a narrow, bright hydrogen line (\ion{H}{1} 1.876 \mic) confirm the value for spectral resolution. In the 3 \mic\ channel, the effective spectral resolution is broadened up to 50\% by the high-frequency wavelength jitter described in Section \ref{sec:wavecal}.

\subsection{Data Description}
\begin{figure}[htp]
	\centering
	\includegraphics[width=0.95\linewidth]{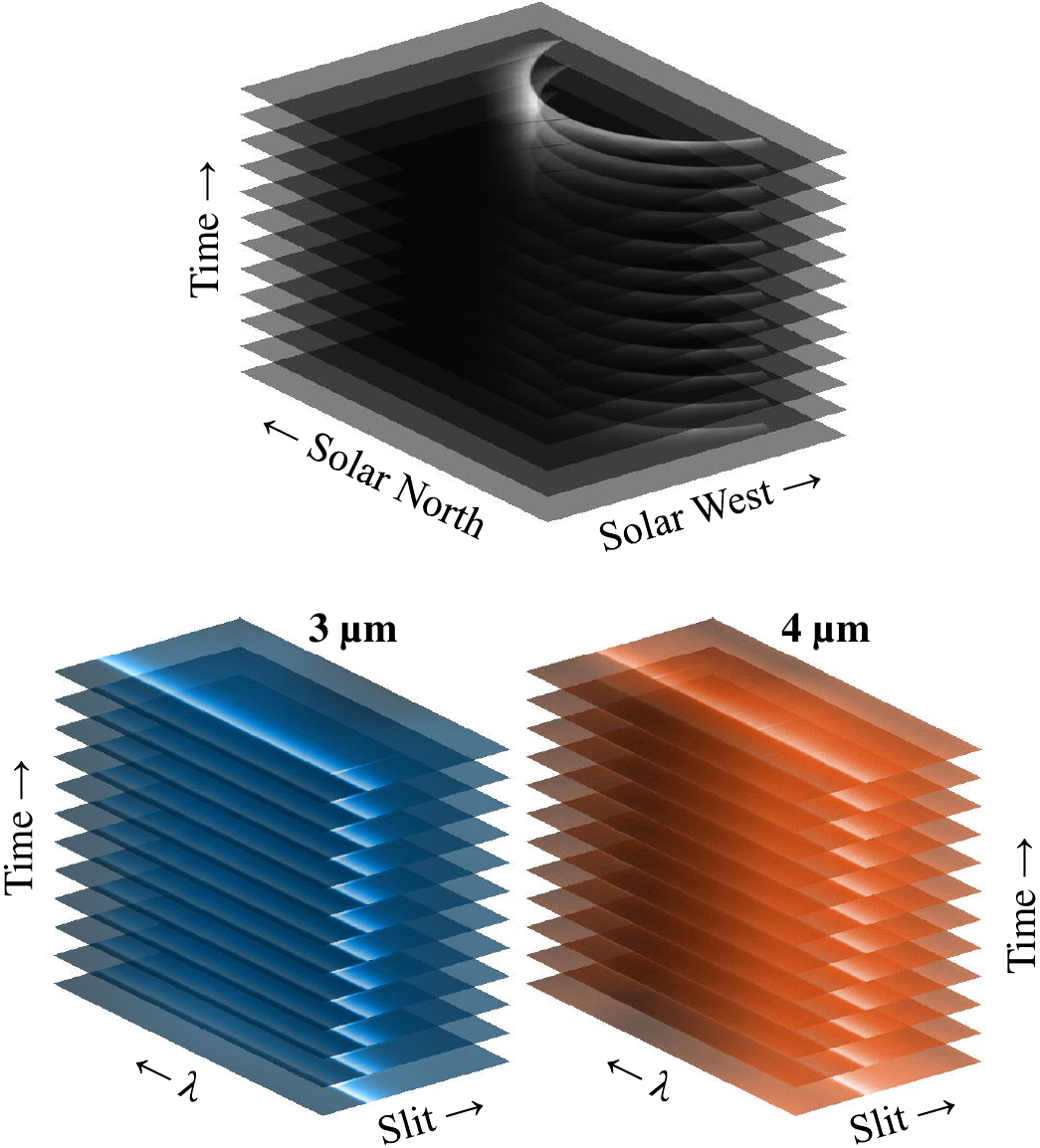}
	\caption{Illustration of data from one of the five eclipse observations. The data set consists of three 3D arrays from the slit-jaw camera and both channels of the IR camera.} 
	\label{fig:data}
\end{figure}

The AIR-Spec data have been organized by slit position into five separate observations (Table \ref{tab:obs-summary}) and are available on the Virtual Solar Observatory at \mbox{\url{https://nso.virtualsolar.org/AIR-Spec/}}. Each data set consists of three 3D arrays with images from the slit-jaw camera and 3 \mic\//4 \mic\ channels of the IR camera, as depicted in Figure \ref{fig:data}.

All slit-jaw frames have been aligned with respect to Moon center and rotated to orient solar north up. Each data point is indexed by solar coordinates and time. The solar coordinates are different for every frame because the Moon drifts with respect to the Sun. Solar $x$ is fixed over rows but varies with column and frame. Solar $y$ is fixed over columns but varies with row and frame. Time is fixed over rows and columns but varies with frame. 

All IR frames have undergone dark subtraction, defective pixel replacement, and geometric correction. Frames have been co-aligned spatially so that the limb crossings coincide.  Each data point is indexed by solar coordinates, wavelength, and time. Solar $x$ and $y$ are fixed over spectral pixel but vary with spatial pixel and frame. Wavelength is fixed over spatial pixel but varies with spectral pixel and frame. Time is fixed over spatial and spectral pixel but varies with frame. 

\section{Future Instrument Improvements}
\label{sec:lessons}

\revis{The 2017 AIR-Spec observations suffered from three limitations: (1) high thermal instrument background, including significant spatial structure and temporal variations,  (2) substantial image jitter, and (3) ghost and double-peak spectral artifacts due to stray reflections in the slit-jaw substrate \citep{Samra2019} and insufficient constraints on the focus mirror mounts. We discovered the spectral artifacts and temporal variation of the instrument background during analysis of the 2017 data. We were aware of the thermal background and image stability issues before the eclipse, but the tight schedule did not permit us to improve the instrument in time to make the observation. We made the decision to observe anyway because we were confident that we would measure the strong \ion{Si}{10} line and hopeful that we might measure or place an upper limit on the intensities of the other four target lines. Ahead of the 2019 eclipse, we made significant improvements and corrected all of the above issues. \citet{Samra2021} explain the improvements in detail. }
	
\revis{The camera was returned to the manufacturer for an upgrade that included a new band-limiting filter with 3~\mic\ cutoff (resulting in the loss of the 3.935~\mic\ \ion{Si}{9} line), improved baffling to block light from the warm window mount, and a lower focal plane temperature of 50 K to reduce the dark current. Altogether, this resulted in a 30x reduction in the instrument background during the 2019 eclipse.  By blocking light at 4 \mic, the new filter had the unintended positive consequence of removing the interfernce fringes from the 4 \mic\ channel. The background temporal variation was removed by minimizing the time between the end of one exposure and the start of the next frame (setting the exposure time as close as possible to the inverse frame rate). A 5x reduction in image jitter was achieved by adding real-time closed-loop feedback to the image stabilization system to supplement the feedforward signal from the gyroscope. The closed-loop error signal was computed by fitting a circle to the lunar limb in the slit-jaw camera. The stray slit-jaw reflections were blocked with a metal shield added to the back of the sapphire substrate, and the focus mirror mounts were better constrained, completely eliminating the broad/double-peak lineshape artifacts.}

\acknowledgments 
{AIR-Spec would never have been conceived without Philip Judge of the NCAR High Altitude Observatory (HAO), whose theoretical work laid the foundation for the eclipse observation. Phil provided invaluable guidance as we planned the experiment and interpreted the results. We are very grateful for his help in making AIR-Spec a success. We gratefully acknowledge SAO colleagues Roger Eng, Sam Fedeler, Tom Gauron, Kim Goins, Giora Guth, Stan Kench, Joyce Medaglia, Brian Robertson, and Dave Weaver for their help in creating AIR-Spec.  We thank Louis Lussier and the team at the NCAR Research Aviation Facility (RAF) for enabling the Gulfstream~V flights, Stuart Beaton of NCAR RAF for specifying, purchasing, and sharing information on the sapphire viewport, Steven Tomczyk of NCAR HAO for lending his equipment, optical expertise, and laboratory space, and Giulio Del Zanna, Helen Mason, Serge Koutchmy, and Shadia Habbal, whose expertise in spectroscopy helped us plan the experiment and interpret the data.  Finally, we thank Robert Kurucz of SAO for providing a photospheric spectrum for the wavelength and radiometric calibrations and Brian McLeod and Giovanni Fazio of SAO for sharing their expertise in the design and implementation of infrared systems. The development of AIR-Spec was made possible by a National Science Foundation Major Research Instrumentation grant, AGS-1531549, with support shared by Smithsonian Institution. J. Hannigan is supported by NASA contract NNX17AE38G.}

\bibliographystyle{aasjournal}
\bibliography{instrument_paper.bib}



\end{document}